\documentclass[12pt]{iopart}

\usepackage{graphicx}
\usepackage{iopams}

\def\vep{\varepsilon}
\def\sign{{\rm sign}}

\begin{document}

\title{Quantum spin Hamiltonians for the $SU(2)_k$ WZW model}

\author{Anne E B Nielsen$^1$, J Ignacio Cirac$^1$ and Germ\'an Sierra$^2$}
\address{$^1$ Max-Planck-Institut f\"{u}r Quantenoptik, Hans-Kopfermann-Str.~1, D-85748 Garching, Germany}
\address{$^2$ Instituto de F\'{\i}sica Te\'orica, UAM-CSIC, Madrid, Spain}

\begin{abstract}
We propose to use null vectors in conformal field theories to derive model Hamiltonians of quantum spin chains and corresponding ground state wave function(s). The approach is quite general, and we illustrate it by constructing a family of Hamiltonians whose ground states are the chiral correlators of the $SU(2)_k$ WZW model for integer values of the level $k$. The simplest example corresponds to $k=1$ and is essentially a nonuniform generalization of the Haldane-Shastry model with long-range exchange couplings. At level $k=2$, we analyze the model for $N$ spin 1 fields. We find that the Renyi entropy and the two-point spin correlator show, respectively, logarithmic growth and algebraic decay. Furthermore, we use the null vectors to derive a set of algebraic, linear equations relating spin correlators within each model. At level $k=1$, these equations allow us to compute the two-point spin correlators analytically for the finite chain uniform Haldane-Shastry model and to obtain numerical results for the nonuniform case and for higher-point spin correlators in a very simple way and without resorting to Monte Carlo techniques.
\end{abstract}


\vspace{1pc}

\begin{indented}
\item[]{\it Keywords\/}: Solvable lattice models, Correlation functions, Conformal field theory
\end{indented}

\section{Introduction}

Chains of interacting spins are important models for investigating properties of quantum many-body systems. The fact that the dimension of the Hilbert space grows exponentially with the length of the chains does, however, limit the applicability of brute force numerical computations. This makes analytical insights particularly valuable, both as a direct source of information and as a starting point for constructing and testing approximate numerical methods. Important analytical techniques include the Bethe ansatz \cite{Bethe} and the construction lying behind the AKLT model \cite{AKLT}. The former was introduced by Bethe in 1931 to solve the antiferromagnetic Heisenberg model with nearest-neighbour interactions and has turned out to be a more generally applicable tool to solve particular spin models, and the latter has led to a number of models for which both the Hamiltonian and the corresponding ground state(s) are known. Solvable models with long-range interactions are more rare, but the Haldane-Shastry (HS) model \cite{H88,S88} is an example. In this model, $N$ spin $1/2$ particles are positioned uniformly on a circle, and each spin interacts with all other spins through a two-body exchange interaction whose strength is inversely proportional to the square of the distance between the spins measured along the chord.

In the present paper, we propose a new method to find matching Hamiltonians and ground state wave functions, which naturally leads to models with long-range interactions. The approach, which is presented in detail in section \ref{sec:hamiltonians}, relies on the existence of null vectors in conformal field theories (CFT), and we shall specifically consider the $SU(2)_k$ WZW models, where $k$ is a positive integer known as the level of the algebra. The vacuum expectation value of a product of primary chiral conformal fields is zero if one of the fields is a null vector, and we use this fact and the Ward identity to derive equations of the form $H\psi=0$, where $H$ is a Hermitian and positive semidefinite operator and $\psi$ is a chiral correlator (i.e., the vacuum expectation value of a product of primary chiral fields). Explicit analytical expressions for several of the relevant chiral correlators have been found recently in \cite{as10}.

In addition to deriving new spin models, the null vectors enable us to derive a set of simple algebraic linear equations, which relate different $n$-point spin correlators within each model (i.e., expectation values of the form $\langle\psi|t_{i_1}^{a_1}\ldots t_{i_n}^{a_n}|\psi\rangle$, where $t_{i_j}^{a_j}$ is the $a_j^{\rm th}$ component of the spin operator acting on spin number $i_j$). The approach turns out to be particularly valuable for $k=1$, where it leads to a closed set of equations for the two-point spin correlators, a closed set of equations for the two-point and the four-point spin correlators and so on. As detailed below, this allows us to obtain a number of results for spin correlators either analytically or through a very simple numerical computation.

After explaining the general approach, which is valid for all $k$, we turn to a more detailed investigation of models obtained at levels $k=1$ and $k=2$ in sections \ref{sec:keq1} and \ref{sec:keq2}, respectively. In the $SU(2)_1$ WZW model, there are only spin 0 and spin 1/2 chiral conformal fields, and since the fusion properties of the spin 0 field are trivial, we find the explicit expression for a Hamiltonian which has the chiral correlator of $N$ spin $1/2$ fields as its ground state in section \ref{sec:keq1hw}. The Hamiltonian involves only two-body interactions, and in section \ref{sec:keq1HS}, we show that it is essentially a nonuniform generalization of the HS model with long-range exchange interactions. We also compute the spectrum for a number of choices of distributions for $N=6$ and discuss symmetries and the origin of degeneracies. The Hamiltonian of the nonuniform HS model was put forward recently in \cite{CS10}, where it was shown analytically that the chiral correlator of $N$ spin $1/2$ fields is an eigenstate of it. This was done by using the Knizhnik-Zamolodchikov (KZ) equation \cite{KZ84}, and we provide the details of that derivation in \ref{sec:alternative}. In contrast to the present work, however, this approach cannot be straightforwardly generalized to higher values of $k$, and it does not automatically guarantee that the chiral correlator is, indeed, the ground state.

The nonuniform HS model was compared to other spin models through exact numerical computations for small $N$ in \cite{CS10}, and the finite chain Renyi entropy and the two-point spin correlation function was studied numerically through Monte Carlo simulations. In the present paper, we instead concentrate on the analytical and semi-analytical expressions for the spin correlators that follow from the properties of the null vectors through the aforementioned linear equations. We rederive the expression for the two-point spin correlator of the uniform HS model in the thermodynamic limit and furthermore demonstrate that the two-point spin correlator can also be computed analytically for the finite chain uniform HS model. These results are presented in section \ref{sec:keq1CF}, where we also provide further numerical results for two-point and four-point spin correlators.

Section \ref{sec:keq2} is devoted to an investigation of the case $k=2$. In this model, there are spin 0, spin 1/2 and spin 1 chiral conformal fields, and we shall, in particular, analyze the spin chain model in which the wave function is the chiral correlator of a product of $N$ spin 1 fields. In section \ref{sec:Ham1}, we derive the Hamiltonian and discuss the spectrum. In section \ref{sec:bilinear}, we find that the wave function has an overlap with the ground state of the bilinear-biquadratic spin 1 chain close to unity in a region within the Haldane phase for $N=6$. In section \ref{sec:entcor1}, we compute the Renyi entropy and the two-point spin correlation function by Monte Carlo simulations. We find that the Renyi entropy grows logarithmically with the length of the considered subchain, while the two-point spin correlation function decays algebraically with the separation between the spins. The Hamiltonian for the chiral correlator of a product of $N$ spin 1/2 fields is given in \ref{sec:k2spinhalf} together with a discussion of the spectrum. Finally, concluding remarks are given in section \ref{sec:conclusion}.

\section{Hamiltonians for the $SU(2)_k$ WZW model} \label{sec:hamiltonians}

The WZW models, together with the minimal models of Belavin, Polyakov and Zamolodchikov, are the best known examples of rational conformal field theories (RCFT). In a RCFT the number of primary fields is finite, and they form a closed operator algebra characterized by the fusion rules.
In this paper, we shall restrict ourselves to the WZW model based on the Kac-Moody (KM) algebra $SU(2)_k$. There are $k+1$ primary fields in this RCFT, which are denoted by $\phi_j$, where $j=0, \frac{1}{2}, \dots, \frac{k}{2}$ labels the total spin. Each field consists of $2j+1$ components corresponding to the $2j+1$ possible values of the projection of the spin on the quantization axis. To refer explicitly to one of these components, we use the notation $\phi_{jm}$, where $m=-j,-j+1,\ldots,j$. The Virasoro central charge $c$, conformal weights $h_j$ and fusion rules are given by
\begin{equation}\label{su2}
c=\frac{3k}{k+2}, \qquad h_j = \frac{j(j+1)}{k+2}, \qquad \phi_{j_1} \times \phi_{j_2} = \sum_{j= |j_1-j_2|}^{ {\rm min} (j_1+j_2, k - j_1-j_2)} \phi_j.
\end{equation}
The simplest nontrivial WZW model corresponds to $k=1$ and has $c=1$ and two primary fields $\phi_0$ and $\phi_{1/2}$ with $h_0=0$ and $h_{1/2} = 1/4$. This model can be easily constructed using a free massless boson whose central charge is $c=1$. The next model in this series corresponds to $k=2$ and has $c = 3/2$ and three primary fields $\phi_0$, $\phi_{1/2}$ and $\phi_1$ with $h_0=0$, $h_{1/2} = 3/16$ and $h_1 = 1/2$. This model can be constructed by combining three Ising models ($c = 3 \times \frac{1}{2}$), or alternatively it can be built using a free boson and an Ising model ($c = 1 + \frac{1}{2}$). In this paper, we shall focus on the $k=1$ and $2$ models, but the general formalism will be valid for any integer level $k \geq 1$.

\subsection{Null vectors}

The {\em rationality} of the WZW models lies in the existence of null vectors in the representation spaces of KM algebras. The decoupling of these vectors provides the basic equations that allow for a complete solution of these models. Null vectors will be the key for the construction of Hamiltonians associated to WZW models, and for that reason we shall describe them in detail below.

The $SU(2)_k$ WZW model is characterized by the chiral currents $J^a(z) \; (a=1,2,3)$, whose Laurent expansion $J^a(z) = \sum_{n = - \infty}^\infty J^a_n \, z^{ -n -1}$ contains the modes $J^a_n$ that satisfy the KM algebra
\begin{equation}\label{km}
[ J_n^a, J_m^b] = \rmi \sum_c \; \vep_{abc} \; J^c_{n+m} + \frac{k}{2} n \; \delta_{a b} \, \delta_{n+m,0}, \qquad n,m \in \mathbb{Z},
\end{equation}
where $\vep_{abc}$ is the Levi-Civita symbol. Notice that the zero modes $J^a_0$ form a closed $SU(2)$ algebra. In the standard spin basis the algebra becomes
\begin{eqnarray}\label{km2}
\eqalign{[ J_n^0, J_m^0] = \frac{k}{2} n \, \delta_{n+m,0}, \\
\left[ J_n^0, J_m^{\pm} \right] = \pm  J^{\pm}_{n+m}, \\
\left[ J_n^+ , J_m^- \right] = 2 J^0_{n+m} + k \, n \, \delta_{n+m,0},}
\end{eqnarray}
where $J^0_n = J^3_n, \, J^\pm_n = J^1_n \pm \rmi J^2_n$.

To construct the representation theory of this KM algebra, we first note that the primary fields introduced above effectively factorize into a product of a chiral field and an anti-chiral field. We shall only consider the former, which we denote by $\phi_j(z)$, where $z$ is a complex parameter. In the context of conformal field theory, $z$ represents a point in two-dimensional space-time, but later on, these coordinates will turn into free parameters in the Hamiltonians we construct. To each chiral primary field $\phi_j(z)$, one associates a primary state $| \phi_j \rangle$ satisfying the following conditions
\begin{equation}\label{prim}
|\phi_j \rangle = \phi_j(0) \, | 0 \rangle, \qquad J^a_0 \, |\phi_j \rangle = t^a |\phi_j \rangle, \qquad J^a_n \, |\phi_j \rangle =0, \qquad n > 0,
\end{equation}
where $|0\rangle$ is the vacuum of the theory (i.e., $J^a_{n \geq 0} |0 \rangle =0$) and $t^a$ are the $2j+1$ dimensional matrices of the
spin $j$ representation of $SU(2)$ (the $m$th component of $t^a |\phi_j \rangle$ is $\sum_{m'}t_{mm'}^a |\phi_{jm'} \rangle$).

The representation spaces of the KM algebra are modules ${\cal H}_j$  obtained by acting on the primary states with products of the negative mode operators, i.e.\ $J^{a_1}_{- n_1 } J^{a_2}_{- n_2 } \dots J^{a_M}_{- n_M } |\phi_j \rangle \; ( n_1 , n_2, \dots ,n_M \geq 1)$. These states are called descendants and can be organized in towers having a common value of the sum $n= n_1 + \dots + n_M$ that is called the Virasoro level, not to be confused with the level $k$ of the KM algebra.

A null vector $|\chi \rangle$ is a descendant state which is also primary, i.e., satisfies the condition (recall (\ref{prim}))
\begin{equation}
J^a_n \, |\chi \rangle = 0, \qquad n > 0.
\label{null}
\end{equation}
Each module ${\cal H}_j$ contains infinitely many null vectors. Here, we shall consider the null vector \cite{GW}
\begin{equation}
| \chi_{j_* j_*} \rangle  = \left( J^+_{-1} \right)^{n_*} \; | \phi_{jj} \rangle,  \qquad n_* = k+1 - 2 j,  \qquad j_* = k+1 - j,
\label{gw}
\end{equation}
which is the null vector occurring at the lowest Virasoro level. It is illustrative to verify (\ref{null}) for the case $J ^a_n = J^-_1$. Using the commutator
\begin{equation}
[ J^-_1, \left( J^+_{-1} \right)^\ell ] = \ell \left( J^+_{-1} \right)^{\ell -1} \left( k +1 - \ell - 2 J^0_0 \right), \qquad \ell \geq 1,
\end{equation}
that follows from (\ref{km2}) by induction, one obtains,
\begin{equation}
\eqalign{J^-_1 | \chi_{j_* j_*} \rangle &= [ J^-_1, \left( J^+_{-1} \right)^{k+1 - 2 j }] \; | \phi_{jj} \rangle \\
&=( k + 1 - 2 j) \left( J^+_{-1} \right)^{k - 2 j } ( 2 j - 2 J^0_0) | \phi_{jj} \rangle = 0,}
\end{equation}
where we have used that $| \phi_{j j} \rangle$ is a primary state,
$J^-_1 | \phi_{j j} \rangle=0$, and that its spin projection is $j$, $J^0_0 | \phi_{j j} \rangle=j | \phi_{j j} \rangle$.
One can easily prove that
\begin{equation}
J^+_0 | \chi_{j_* j_*} \rangle = 0, \qquad J^0_0 | \chi_{j_* j_*} \rangle = j_*  | \chi_{j_* j_*} \rangle,
\end{equation}
so that $ | \chi_{j_* j_*} \rangle $ is the highest weight vector of a representation with spin $j_*$. In summary, the module ${\cal H}_j$ contains at level $n_*$ a multiplet of $2j_*+1$ null vectors with spin $j_*$.

Since a null vector is also a primary field, a null vector and its descendants comprise a module on their own, and this module can be shown to decouple from ${\cal H}_j$ \cite{cft-book}. The number of states $d_{j,k}(n)$ appearing at Virasoro level $n$ after removing all the decoupled modules can be determined by considering the character
\begin{equation}
\chi_j (q,k) = {\rm Tr}_{{\cal H}_j}  \, q^{ L_0 - c_k/ 24 }=
q^{h_j - c_k/24} \sum_{n=0}^\infty d_{j, k}(n) \; q^n.
\end{equation}
This character is given by \cite{cft-book}
\begin{equation}
 \chi_j (q,k) = \frac{ q^{ (2 j +1)^2/[4(k+2)]}}{ \eta(q)^3}
 \sum_{n = - \infty}^\infty  [ 2 j + 1 + 2 n (k+2) ]\; q^{ n [ 2 j + 1 + n (k+2)]},
\end{equation}
where $\eta(q)$ is the Dedekin eta function
\begin{equation}
\eta(q) = q^{1/24} \prod_{n=1}^\infty (1- q^n).
\end{equation}
We collect the values of $d_{j,k}(n)$ for the lowest spin modules ${\cal H}_j$ and various values of $n$ and $k$ in table \ref{table1}. An inspection of the rows of this table shows that, for given $n$, the number of states remains constant beyond a certain value of $k$. This can be understood from the fact that states are removed whenever there is a null vector. For $k\leq n^*-1+2j$ (recall (\ref{gw})), one or more null vectors are present for $n\leq n^*$, whereas no null vectors are present for $n\leq n^*$ if $k>n^*-1+2j$. The number of states removed at $k= n^*-1+2j$ and $n=n^*$ is the number of null vectors in the multiplet $|\chi_{j_* m_*}\rangle$, i.e. $2j_*+1$. This is expressed in the relation
\begin{equation}
d_{j,k}(n_*) - d_{j,n^*-1+2j}(n_*) = 2 j_* +1, \qquad \forall k > n^*-1+2j.
\label{8}
\end{equation}
In the module ${\cal H}_0$, for example, there are $4$ states at $n=2$ and $k=1$, which turn into $9$ states for all $k > 1$. The difference $9-4=5$ is the number of null vectors that form the $j_*=2$ spin multiplet at level $n_*=2$ and $k=1$.

\begin{table}
\caption{\label{table1}Number of states $d_{j,k}(n)$ for the modules ${\cal H}_j \; \, (j=0, \frac{1}{2}, 1)$. The value of $d_{j,k}(n_*)$ appears with a star. The Virasoro level $n$ labels the rows, and the KM level $k$ labels the columns.}
\vspace{1mm}
\scalebox{0.81}{
\begin{tabular}{| c | c c c c c c | c c c c c c | c c c c c c |}
\hline
&\multicolumn{6}{c |}{${\cal H}_0$}&\multicolumn{6}{c |}{${\cal H}_{1/2}$} & \multicolumn{6}{c |}{${\cal H}_{1}$} \\
\cline{2-19}
$n | k$ &1 & 2 & 3 & 4 & 5 & \dots & 1 & 2 & 3 & 4 & 5& \dots & 2 & 3 & 4 & 5 & 6 & \dots \\
\hline
0 & 1 & 1 & 1  &1  &1  & 1 & 2 & 2& 2 &2  & 2 & 2 & 3 & 3 & 3 & 3 & 3& 3 \\
1 & 3& 3 & 3  & 3 & 3 & 3 & ${\bf 2^*}$ & 6 &6  &6  &6  &6  & ${\bf 4^*}$  & 9 & 9 & 9 & 9 & 9 \\
2 & ${\bf 4^*}$ & 9 & 9  & 9 & 9 & 9 &6 & ${\bf 12^*}$  & 18 &18  & 18  & 18 & 12 & ${\bf 20^*}$   &27  & 27 & 27 & 27 \\
3 & 7 & ${\bf 15^*}$  & 22  &22  &22  &22 & 8 & 26 &${\bf 36^*}$   &44  & 44  & 44  & 21  & 45  & ${\bf 57^*}$   & 66 & 66 & 66 \\
4 & 13 & 30 & ${\bf 42^*}$   & 51 & 51 & 51 & 14 & 48 & 78 & ${\bf 92^*}$  &102  & 102 & 43 & 90  & 126  & ${\bf 142^*}$  & 153 & 153 \\
\hline
\end{tabular}}
\end{table}

\subsection{Clebsch-Gordan coefficients and projectors}

Equation (\ref{gw}) corresponds to the Clebsch-Gordan (CG) decomposition
\begin{equation}
V_{j_*} \subset V_1^{ \otimes  n_*} \otimes V_j
\label{cg1}
\end{equation}
where $V_J = \mathbb{C}^{\otimes (2 J+1)}  \; (J=j_*, 1, j)$ is the vector space for the spin $J$ irreducible representation (irrep). This decomposition consists in the symmetrized product of $n_*$ copies of the spin 1 irrep tensored with the spin $j$ irrep to yield the highest possible spin $j_*= n_* + j$. A basis of $V_{j_*}$ is given by ($m_*=-j_*,-j_*+1,\ldots,j_*$)
\begin{equation}
|\chi_{j_* m_*}\rangle = \sum_{a_1, \dots, a_{n_*}, m}   C^{j_* m_*}_{ a_1, \dots, a_{n_*}, j m} \; J^{a_1}_{-1} \dots J^{a_{n_*}}_{-1} \,
|\phi_{j m}\rangle.
\label{cg2}
\end{equation}
The constants $C^{j_* m_*}_{ a_1, \dots, a_{n_*}, j m}$ can be constructed from the product of $n_* $ CG coefficients and satisfy the condition
\begin{equation}
\sum_{a_1, \dots, a_{n_*}, m}  C^{j_* m_*}_{ a_1, \dots, a_{n_*},j m} \, C^{j_* m'_*}_{ a_1, \dots, a_{n_*},j m} = \delta_{m_* m'_*},
\label{cg3}
\end{equation}
which ensures the orthonormality of the vectors $|\chi_{j_* m_*}\rangle$. We adopt the Condon-Shortley convention that all CG coefficients are real. We note that (\ref{cg2}) could equally well be written in terms of the fields $\chi_{j_*m_*}(z)$ and $\phi_{jm}(z)$, and we use this notation in the following.

Let us next define the linear map
\begin{equation}\label{cg4}
\eqalign{{\bf C}_{n_*,j}^{j_*} : \; V_1^{ \otimes n_*} \otimes V_j \rightarrow V_{j_*},\\
C^{j_* m_*}_{ a_1, \dots, a_{n_*},j m} = \langle m_* | {\bf C}_{n_*,j}^{j_*} | a_1, \dots, a_{n_*},m \rangle,}
\end{equation}
which allows us to write (\ref{cg3}) as
\begin{equation}
{\bf C}_{n_*,j}^{j_*} \, \left( {\bf C}_{n_*,j}^{j_*} \right)^\dagger = {\bf I}_{2 j_*+1}.
\label{cg5}
\end{equation}
where ${\bf I}_{2 j_*+1}$ is the identity matrix in $V_{j_*}$. Multiplying ${\bf C}_{n_*,j}^{j_*}$ and $({\bf C}_{n_*,j}^{j_*})^\dagger$ in reverse order
\begin{eqnarray}
{\bf K}_{n_*,j}  =\left( {\bf C}_{n_*,j}^{j_*} \right)^\dagger  {\bf C}_{n_*,j}^{j_*}
\label{cg6}
\end{eqnarray}
yields the map
\begin{equation}
{\bf K}_{n_*,j} : \; V_1^{ \otimes n_*} \otimes V_j \rightarrow V_1^{ \otimes n_*} \otimes V_j
\label{cg7}
\end{equation}
that, using (\ref{cg5}) and (\ref{cg6}),  satisfies
\begin{equation}
 {\bf K}_{n_*,j}^2 = {\bf K}_{n_*,j} , \qquad  {\bf K}_{n_*,j}^\dagger  = {\bf K}_{n_*,j},
\label{cg8}
\end{equation}
so that ${\bf K}_{n_*,j}$ is a projector with components
\begin{equation}
\left( {\bf K}_{n_*, j}  \right)^{ a_1, \dots, a_{n_*},m}_{ b_1, \dots, b_{n_*},m'} =
\sum_{ m_*}  C^{j_* m_*}_{ a_1, \dots, a_{n_*},j m} \, C^{j_* m_*}_{ b_1, \dots, b_{n_*},j m'}.
\label{cg9}
\end{equation}
Using this equation, one can parameterize the null vectors (\ref{cg2}) with indices of the vector space $V_1^{ \otimes n_*} \otimes V_j$ as follows
\begin{equation}
\eqalign{\chi_{a_1, \dots, a_{n_*}, j m}(z) & \equiv \sum_{m_*} C^{j_* m_*}_{ a_1, \dots, a_{n_*},j m} \; \chi_{j_* m_*}(z)\\
& = \sum_{b_1, \dots, b_{n_*}, m'}  \left( {\bf K}_{n_*, j}  \right)^{ a_1, \dots, a_{n_*},m}_{ b_1, \dots, b_{n_*},m'} \; \;
J^{b_1}_{-1} \dots J^{b_{n_*}}_{-1} \, \phi_{j m'}(z).}
\label{cg10}
\end{equation}
Since, by use of (\ref{cg3}),
\begin{equation}
\chi_{j_* m_*}(z) = \sum_{a_1, \dots, a_{n_*}, m} C^{j_* m_*}_{ a_1, \dots, a_{n_*},j m} \; \chi_{a_1, \dots, a_{n_*}, j m}(z),
\end{equation}
we see that $\chi_{a_1, \dots, a_{n_*}, j m}(z)$ fulfil the eigenvalue equation
\begin{equation}
\chi_{a_1, \dots, a_{n_*}, j m}(z) = \sum_{b_1, \dots, b_{n_*}, m' } \left( {\bf K}_{n_*, j}  \right)^{ a_1, \dots, a_{n_*},m}_{ b_1, \dots, b_{n_*},m'} \; \chi_{b_1, \dots, b_{n_*}, j m'}(z).
\label{cg10b}
\end{equation}

As a result of the above construction, ${\bf K}_{n_*, j}$ must fulfil the following symmetry constraints
\begin{eqnarray}
\left( {\bf K}_{n_*, j}  \right)^{\ldots a_i \dots a_j \ldots,m}_{\ldots b_k \dots b_l \ldots,m'}=\left( {\bf K}_{n_*, j}  \right)^{\ldots a_j \dots a_i \ldots,m}_{\ldots b_k \dots b_l \ldots,m'}=
\left( {\bf K}_{n_*, j}  \right)^{\ldots a_i \dots a_j \ldots,m}_{\ldots b_l \dots b_k \ldots,m'},\label{cg10c1}\\
\sum_{a_i,a_j}\delta_{a_ia_j}\left( {\bf K}_{n_*, j}  \right)^{\ldots a_i \dots a_j \ldots,m}_{ b_1, \dots, b_{n_*},m'}=0, \qquad
\sum_{b_i,b_j}\left( {\bf K}_{n_*, j}  \right)^{a_1, \dots, a_{n_*},m}_{\ldots b_i \dots b_j \ldots,m'}\delta_{b_ib_j}=0,\label{cg10c2}\\
\sum_{a_i,m''}t^{a_i}_{mm''}\left( {\bf K}_{n_*, j}  \right)^{\ldots a_i \dots,m''}_{ b_1, \dots, b_{n_*},m'}=0, \qquad
\sum_{b_i,m'}\left( {\bf K}_{n_*, j}  \right)^{a_1, \dots, a_{n_*},m}_{\ldots b_i \dots,m'}t^{b_i}_{m'm''}=0,\label{cg10c3}
\end{eqnarray}
and, indeed, (\ref{cg10c1}-\ref{cg10c3}) together with (\ref{cg8}) are sufficient to uniquely determine ${\bf K}_{n_*, j}$. The null vectors at levels $k=1$ and $2$ are collected in table \ref{table2}, and one can check that the following $K$-tensors fulfil the conditions,
\begin{eqnarray}
\fl\left( {\bf K}_{ 2 , 0}  \right)^{a_1 a_2}_{ b_1 b_2} = \frac{1}{2} ( \delta_{a_1 b_1}
\delta_{a_2 b_2} + \delta_{a_1 b_2} \delta_{a_2 b_1} ) - \frac{1}{3} \delta_{a_1 a_2} \delta_{b_1 b_2}, \label{cg12} \\
\fl\left( {\bf K}_{ 1 , \frac{1}{2} } \right)^{a m}_{ b  m' } = \frac{2}{3} \left( \delta_{ ab} \, \delta_{ m m'} - \rmi \sum_c \vep_{abc} t^c_{m m'} \right), \qquad t^a = \frac{1}{2} \sigma^a, \label{cg13} \\
\eqalign{\fl\label{cg14}\left( {\bf K}_{ 3 , 0}  \right)^{a_1 a_2 a_3}_{ b_1 b_2 b_3 } = \frac{1}{6} \left( \delta_{a_1 b_1} \delta_{a_2 b_2}  \delta_{a_3 b_3} +\; {\rm permutations} \; {\rm of} \; b' {\rm s} \right)
-\frac{1}{15} \big(\delta_{a_1 a_2} \delta_{b_1 b_2}  \delta_{a_3 b_3}\\
+ \delta_{a_2 a_3} \delta_{b_2 b_3}  \delta_{a_1 b_1}+ \delta_{a_1 a_3} \delta_{b_1 b_3}  \delta_{a_2 b_2}
+\; {\rm cyclic} \;   {\rm permutations} \; {\rm of} \; b' {\rm s}\big),}\\
\eqalign{\fl\label{cg15}\left( {\bf K}_{ 2 , \frac{1}{2}}  \right)^{a_1 a_2 m }_{ b_1 b_2 m' } = \left[  \frac{3}{10} ( \delta_{a_1 b_1}
\delta_{a_2 b_2} +  \delta_{a_1 b_2} \delta_{a_2 b_1} ) - \frac{1}{5} \delta_{a_1 a_2} \delta_{b_1 b_2}  \right] \delta_{m m'}\\
-\frac{\rmi}{5} \sum_c \left( \vep_{a_1 b_1 c} \, \delta_{a_2 b_2}
+\vep_{a_2 b_2 c} \, \delta_{a_1 b_1} + \vep_{a_1 b_2 c} \, \delta_{a_2 b_1}
+\vep_{a_2 b_1 c} \, \delta_{a_1 b_2} \right) \, t^c_{m m'},}\\
\fl\left( {\bf K}_{ 1 , 1} \right)^{a m}_{ b m'} = \frac{1}{2} ( \delta_{a b}
\delta_{m m'} +  \delta_{a m'} \delta_{m b} ) - \frac{1}{3} \delta_{a m} \delta_{b m'}, \label{cg16}
\end{eqnarray}
where $\sigma^a$ ($a=1,2,3$) are the Pauli matrices. To simplify the notation, we shall drop the $m$ and $m'$ indices in the following, considering that $K$ is a matrix, and we shall let the summation over repeated spin component indices (e.g., $b_1,\ldots,b_{n_*}$) be implicit.

\begin{table}
\caption{\label{table2}Null vectors for $k=1$ and $2$. The possible values of the indices are: $a,a_1,a_2,a_3=1,2,3$, $m= \pm 1/2$, and $m'=0,\pm 1$.}
\begin{indented}
\item[]\begin{tabular}{@{}lllll}
\br
$k$ & $j$ & $n_*$ &  $j_*$ & $\chi^{a_1, \dots, a_{n_*}, j m}$ \\
\mr
1 & 0& 2&  2 & $\chi^{a_1 a_2, 0 0}$ \\
1 &  $\frac{1}{2}$& 1  & $\frac{3}{2}$ & $ \chi^{a,\frac{1}{2} m}$  \\
\mr
2 & 0& 3 & 3 & $\chi^{a_1 a_2 a_3, 0 0}$\\
2 & $\frac{1}{2}$& 2  & $\frac{5}{2}$ & $\chi^{a_1 a_2, \frac{1}{2} m} $ \\
2 & 1& 1 & 2  & $\chi^{a, 1 m'}$ \\
\br
\end{tabular}
\end{indented}
\end{table}

\subsection{Decoupling equations}

A chiral correlator is the vacuum expectation value of the product of a set of primary chiral fields:
\begin{equation}
\psi(z_1, \dots, z_N) = \langle \phi_{j_1}(z_1) \dots
\phi_{j_N}(z_N) \rangle.
\label{dec1}
\end{equation}
In our construction, $\psi(z_1, \dots, z_N)$ is the wave function. Written more explicitly, the state of the spin chain is
\begin{equation}
|\psi(z_1, \dots, z_N)\rangle=\sum_{m_1,\ldots,m_N} \psi_{m_1,\ldots,m_N}(z_1,\ldots,z_N)|m_1,\ldots,m_N\rangle,
\end{equation}
where the coefficients are given by
\begin{equation}
\psi_{m_1\ldots m_N}(z_1, \dots, z_N) = \langle \phi_{j_1m_1}(z_1) \dots
\phi_{j_Nm_N}(z_N) \rangle.
\end{equation}
We note that a chiral correlator is not necessarily one unique function. If the fusion rules (\ref{su2}) allow more than one possibility for how two fields can fuse (i.e., if there is more than one field on the right hand side of (\ref{su2})), there will be more different fusion channels, and each fusion channel corresponds to one wave function. All of these wave functions are ground states of the Hamiltonian that we shall construct.

There are more approaches available to find explicit expressions for the chiral correlators (\ref{dec1}). In some simple cases it is possible to solve the KZ equations \cite{KZ84}, which are a system of partial differential equations in the coordinates $z_i$ fulfilled by the chiral correlators. The KZ equations can be derived from the null vector $\chi = (L_{-1} - \frac{ k + 2}{2} J^a_{-1} J^a_0) \phi$, where $L_{-1}$ is one of the Virasoro generators. Another possibility is to use the Feigin-Fucks or Coulomb gas representation of the WZW model \cite{cft-book}. The aim of this section is to derive a set of algebraic equations satisfied by the chiral correlators that follow from the decoupling of null vectors in the modules ${\cal H}_j$. These equations are algebraic, and not differential, as in the KZ case, and this fact will allow us to construct quantum spin Hamiltonians whose ground states are the chiral correlators of the WZW model.

Before presenting this construction, let us notice that (\ref{dec1}) is invariant under global $SU(2)$ rotations generated by the total spin operator $T^a$  \cite{GW}
\begin{equation}
T^a \, \psi = \sum_{i=1}^N t^a_i \; \langle \phi_{j_1}(z_1) \dots \phi_{j_N}(z_N) \rangle =0, \qquad T^a \equiv \sum_{i=1}^N t^a_i,
\label{dec2}
\end{equation}
where $t^a_i$ denotes the spin matrix acting at the site $i$, in the representation $j_i$ corresponding to the field $\phi_{j_i}(z_i)$. We shall suppose for simplicity that $j_i = j$, $\forall i$, but the results can be easily generalized when the primary fields have different spins.

Let us start by replacing the primary field $\phi_j(z_i)$ in the correlator (\ref{dec1}) by the null field (\ref{cg10}) in the module ${\cal H}_{j}$. One can show from the definition of a null field that a chiral correlator is zero if it contains a null field \cite{cft-book}. This is referred to as the decoupling of the null field. Therefore,
\begin{equation}
 \langle \phi_j(z_1) \dots   \chi^{a_1, \dots, a_{n_*} j}(z_i)  \dots \phi_j(z_N) \rangle = 0,
\label{dec3}
\end{equation}
i.e.,
\begin{equation}
\left( {\bf K}_{n_*, j}^{(i)}  \right)^{ a_1, \dots, a_{n_*}}_{ b_1, \dots, b_{n_*}}
 \langle \phi_j(z_1) \dots   (J^{b_1}_{-1} \dots J^{b_{n_*}}_{-1} \, \phi_j)(z_i)  \dots \phi_j(z_N) \rangle = 0
\label{dec4}
\end{equation}
for all $z_1, \dots, z_N$, where the superindex $i$ in ${\bf K}_{n_*, j}^{(i)}$ implies that this matrix acts on the spin degrees of the operator $\phi_j$ representing the $i$th spin in the chain. The modes $J_{-1}^b$ act on the field located at the point $z_i$, but their action can be transferred to the primary fields located at the other positions by means of the Ward identity \cite{GW}
\begin{equation}
\fl\langle \phi_j(z_1) \dots  (J_{-1}^a \psi)(z_i) \dots  \phi_j(z_N) \rangle =\sum_{i_1 (\neq i)}^N \frac{ t^a_{i_1}}{z_i- z_{i_1}} \langle \phi_j(z_1) \dots \psi(z_i) \dots \phi_j(z_N) \rangle.
\label{dec5}
\end{equation}
Here, and in all that follows, we use the convention that the sum over $i_1\neq i_2\neq \ldots\neq i_n(\neq i_{n+1}, \ldots, i_{n+m})$ means that the indices outside (inside) the brackets should (should not) be summed over, and only terms for which all $n+m$ indices are different contribute to the sum. Similarly, we write $i_1\neq i_2\neq \ldots\neq i_n$ when all $n$ indices are required to be different. Applying (\ref{dec5}) $n_*$ times to (\ref{dec4}) yields
\begin{equation}
\sum_{i_1 (\neq i)}^N \dots \sum_{i_{n_*} (\neq i)}^N
 \frac{ \left( {\bf K}_{n_*, j}^{(i)}  \right)^{a_1 \dots a_{n_*}}_{ b_1 \dots b_{n_*}} \; t^{b_1}_{i_1} \dots t^{b_{n_*}}_{i_{n_*}}  }{
 ( z_i- z_{i_1}) \dots ( z_i- z_{i_{n_*}})} \,  \langle \phi_j(z_1) \dots \phi_j(z_N) \rangle = 0,
\label{dec6}
\end{equation}
which can be written in the compact form
\begin{equation}
 {\cal P}_{n_*, j}^{i, a_1 \dots a_{n_*}} ( z_1, \dots, z_N) \, \psi(z_1, \dots, z_N) = 0, \qquad i=1, \dots, N,
 \label{dec7}
 \end{equation}
where
\begin{equation}
 {\cal P}_{n_*, j}^{i, a_1 \dots a_{n_*}}  ( z_1, \dots, z_N) = \sum_{i_1, \dots, i_{n_*} (\neq i)}^N
\left(  {\bf K}_{n_*, j}^{(i)}  \right)^{a_1 \dots a_{n_*}}_{ b_1 \dots b_{n_*}} z^{-1}_{ i i_1} \dots z^{-1}_{ i i_{n_*}} t^{b_1}_{i_1} \dots t^{b_{n_*}}_{i_{n_*}}
\label{dec8}
\end{equation}
and $z_{i k} \equiv z_i - z_k$. These are the algebraic equations mentioned at the beginning of this subsection.

Finally, we shall transform (\ref{dec7}) into the following equation
\begin{equation}
{\cal C}_{n_*, j}^{i, a_1 \dots a_{n_*}} ( z_1, \dots, z_N) \, \psi(z_1, \dots, z_N) = 0, \qquad i=1, \dots, N,
\label{dec9}
\end{equation}
where
\begin{equation}
{\cal C}_{n_*, j}^{i, a_1 \dots a_{n_*}}  ( z_1, \dots, z_N)
= \sum_{i_1, \dots, i_{n_*} (\neq i)}^N
\left(  {\bf K}_{n_*, j}^{(i)}  \right)^{a_1 \dots a_{n_*}}_{ b_1 \dots b_{n_*}} w_{ i i_1} \dots w_{ i i_{n_*}} t^{b_1}_{i_1} \dots t^{b_{n_*}}_{i_{n_*}}
\label{dec10}
\end{equation}
and $w_{i k} \equiv (z_i+z_k)/(z_i-z_k)$. To show this relation, let us apply the identity
\begin{equation}
\frac{2 z_i}{ z_i - z_k} = 1 + \frac{ z_i + z_k}{ z_i - z_k} = 1 + w_{ik}
\end{equation}
to
\begin{eqnarray}
\eqalign{\fl( 2 z_i)^{n_*}  \;  {\cal P}_{n_*, j}^{i, a_1 \dots a_{n_*}} \, \psi =
\left(  {\bf K}_{n_*, j}^{(i)}  \right)^{a_1 \dots a_{n_*}}_{ b_1 \dots b_{n_*}} \sum_{i_1 (\neq i)} \frac{ 2 z_i}{ z_{ i i _1}} t^{b_1}_{i_1} \cdots
 \sum_{i_{n_*} (\neq i)} \frac{ 2 z_i}{ z_{ i i _{n_*}}} t^{b_{n_*}}_{i_{n_*}} \psi
= \left(  {\bf K}_{n_*, j}^{(i)}  \right)^{a_1 \dots a_{n_*}}_{ b_1 \dots b_{n_*}} \\
\times\left( T^{b_1} - t^{b_1}_i + \sum_{i_1 (\neq i)} w_{i i_1} t^{b_1}_{i_1} \right)\cdots
\left( T^{b_{n_*}} - t^{b_{n_*}}_i + \sum_{i_{n_*} (\neq i)} w_{i i_{n_*}} t^{b_{n_*}}_{i_{n_*}} \right) \psi \\
= \left(  {\bf K}_{n_*, j}^{(i)}  \right)^{a_1 \dots a_{n_*}}_{ b_1 \dots b_{n_*}} \left( T^{b_1} + \sum_{i_1 (\neq i)} w_{i i_1} t^{b_1}_{i_1} \right)
 \cdots\left( T^{b_{n_*}} + \sum_{i_{n_*} (\neq i)} w_{i i_{n_*}} t^{b_{n_*}}_{i_{n_*}} \right) \psi }
\end{eqnarray}
where we have used (\ref{cg10c3}) in the final step. Next, we want to move the $T^b$ matrices to the right in this expression. This produces extra terms coming from the commutator
\begin{equation}
[ T^b, \; \sum_{ k (\neq i)} w_{i k} \, t^{b_k}_k ] = \rmi \vep_{ b b_k b'_k} \sum_{ k (\neq i)} w_{i k} \, t^{b'_k}_k,
\end{equation}
but the antisymmetry in the $b$ indices causes these terms to cancel out when multiplied by the tensor $K$ whose indices are totally symmetric. Once the $T^{b_k}$ matrices are brought to the right, they annihilate the correlator $\psi$, which proves that
\begin{equation}
( 2 z_i)^{n_*}  \;  {\cal P}_{n_*, j}^{i, a_1 \dots a_{n_*}} \, \psi
= {\cal C}_{n_*, j}^{i, a_1 \dots a_{n_*}} \, \psi
\end{equation}
and thus (\ref{dec9}) is equivalent to (\ref{dec7}).

\subsection{Spin chain Hamiltonians}

We can use the operators in (\ref{dec10}) to define the matrices
\begin{equation}
 H_{n_*, j}^{(i)} \equiv \left( {\cal C}_{n_*, j}^{i, a_1 \dots a_{n_*}}  ( z_1, \dots, z_N) \right)^\dagger
{\cal C}_{n_*, j}^{i, a_1 \dots a_{n_*}}  ( z_1, \dots, z_N),
\label{dec11}
\end{equation}
which are Hermitian, positive semidefinite and rotationally invariant
\begin{equation}\label{dec12}
\left( H_{n_*, j}^{(i)} \right)^\dagger = H_{n_*, j}^{(i)}, \qquad H_{n_*, j}^{(i)} \geq 0, \qquad [ T^a, H_{n_*, j}^{(i)} ] = 0.
\end{equation}
They annihilate the correlator $\psi$,
\begin{equation}
H_{n_*, j}^{(i)} \, \psi(z_1, \dots, z_N) = 0,
\label{dec13}
\end{equation}
and their sum can be taken as the Hamiltonian of a spin chain with spin $j$ at each site, i.e.,
\begin{equation}
H_{n_*, j}= \sum_{i=1}^N H_{n_*, j}^{(i)}.
\label{dec14}
\end{equation}
It follows immediately that $\psi$ is the ground state of this Hamiltonian. If the fusion rules give rise to several chiral correlators, then $ H_{n_*, j}$ will have several ground states in one-to-one correspondence with the chiral correlators. Finally, we note that Hamiltonians that are translationally invariant by one site can be achieved by choosing the coordinates $z_k$ to be equally spaced on the unit circle, i.e.,
\begin{equation}
z_k = \rme^{2 \pi \rmi k/N} \; (k =1, \dots, N) \Longrightarrow H_{n_*, j}: {\rm translationally} \; \; {\rm invariant}.
\label{dec15}
\end{equation}
We shall refer to this choice as the uniform case.

\subsection{Algebraic equations for spin correlators}\label{sec:gencor}

The decoupling equations (\ref{dec9}) also imply certain relations between the different $n$-point spin correlators. Let us write
\begin{equation}
{\cal C}_{n_*, j}^{i, a_1 \dots a_{n_*}} ( z_1, \dots, z_N) = A_{n_*, j}^{i, a_1 \dots a_{n_*}} ( z_1, \dots, z_N) + Q_{n_*, j}^{i, a_1 \dots a_{n_*}} ( z_1, \dots, z_N)
\end{equation}
as the sum of an anti-Hermitian and a Hermitian operator,
\begin{equation}
A_{n_*, j}^{i, a_1 \dots a_{n_*}}
= - \left(A_{n_*, j}^{i, a_1 \dots a_{n_*}}\right)^\dagger,
\qquad
Q_{n_*, j}^{i, a_1 \dots a_{n_*}}
= \left(Q_{n_*, j}^{i, a_1 \dots a_{n_*}} \right)^\dagger,
\end{equation}
and define
\begin{equation}
T_{i_1\ldots i_n}^{b_1\ldots b_n}=t_{i_1}^{b_1}\ldots t_{i_n}^{b_n}.
\end{equation}
The decoupling equations (\ref{dec9}) imply
\begin{equation}
\langle \psi| T_{i_1\ldots i_n}^{b_1\ldots b_n} \, {\cal C}_{n_*, j}^{i, a_1 \dots a_{n_*}} |\psi \rangle = \langle \psi | \left( {\cal C}_{n_*, j}^{i, a_1 \dots a_{n_*}} \right)^\dagger T_{i_1\ldots i_n}^{b_1\ldots b_n} \, | \psi \rangle = 0,
\end{equation}
and adding and subtracting these equations, we find
\begin{eqnarray}
\langle\psi| \left(\left\{ T_{i_1\ldots i_n}^{b_1\ldots b_n} , A_{n_*, j}^{i, a_1 \dots a_{n_*}} \right\} + \left[ T_{i_1\ldots i_n}^{b_1\ldots b_n} , Q_{n_*, j}^{i, a_1 \dots a_{n_*}} \right]\right) |\psi\rangle = 0, \label{ss20a}  \\
\langle\psi| \left(\left[ T_{i_1\ldots i_n}^{b_1\ldots b_n} , A_{n_*, j}^{i, a_1 \dots a_{n_*}} \right] + \left\{ T_{i_1\ldots i_n}^{b_1\ldots b_n} , Q_{n_*, j}^{i, a_1 \dots a_{n_*}} \right\}\right) |\psi\rangle = 0. \label{ss20b}
\end{eqnarray}
Inserting the actual expressions for $A_{n_*, j}^{i, a_1 \dots a_{n_*}}$ and $Q_{n_*, j}^{i, a_1 \dots a_{n_*}}$ and choosing different values of $n$, (\ref{ss20a}) and (\ref{ss20b}) provide a set of linear equations between different spin correlators.

\section{Results for the $SU(2)$ WZW model at level $k=1$} \label{sec:keq1}

We now turn to a more detailed investigation of the case $k=1$. First, we provide explicit expressions for the wave function and the Hamiltonian for $N$ spin 1/2 fields. We then show that the model is a generalization of the HS model and discuss the degeneracies of the spectrum. As mentioned in the Introduction, the overlap with other spin models, the Renyi entropy and the two-point spin correlator have already been computed numerically in \cite{CS10}, and consequently we here concentrate on deriving analytical and semi-analytical expressions for the spin correlators, which is done in the last part of this section.

\subsection{Wave function and Hamiltonian} \label{sec:keq1hw}

At level $k=1$ there are two primary fields $\phi_0$ and $\phi_{1/2}$ whose fusion rules are
\begin{equation}
\phi_{1/2} \otimes \phi_{1/2} = \phi_0, \qquad \phi_{1/2} \otimes \phi_{0} = \phi_{1/2},  \qquad \phi_{0} \otimes \phi_{0} = \phi_{0}.
\label{fuse1}
\end{equation}
These rules tell us the number of chiral correlators that arise from a product of primary fields on a given geometry, which in our case is the sphere. For $k=1$, the nontrivial correlators are obtained by multiplying $N$ fields $\phi_{1/2}$. Equation (\ref{fuse1}) implies that the fusion of these fields produces the field $\phi_0$ if $N$ is even and $\phi_{1/2}$ if $N$ is odd. Since the vacuum expectation value of $\phi_{1/2}$ is zero, and that of $\phi_0$ is not, $N$ must be even to get a nonvanishing correlator
\begin{equation}
\psi_{s_1, \dots, s_N} (z_1, \dots, z_N) = \langle \phi_{1/2, s_1}(z_1) \dots \phi_{1/2, s_N}(z_N) \rangle.
\label{psi1}
\end{equation}
Here, we have made a slight change of notation compared to the previous section, such that $s_i/2=\pm 1/2$ is the third component of the $i^{\rm th}$ spin. In addition, $\psi_{s_1, \dots, s_N} (z_1, \dots, z_N)$ can only be nonzero for $\sum_{i=1}^Ns_i=0$ because $\psi$ is a singlet. Note that the correlator (\ref{psi1}) is unique since the fusion decomposition only generates one field at a time.

The explicit form of (\ref{psi1}) can easily be obtained from the free field representation of the $SU(2)_1$ WZW model in terms of a free boson $\varphi(z)$ whose two-point correlator is $ \langle \varphi(z_1) \varphi(z_2) \rangle = - \ln(z_1 - z_2)$. In the Cartan-Weyl representation, the current algebra operators $J^a(z)$ are given by
\begin{equation}
J^0(z) = \frac{\rmi}{\sqrt{2}} \partial_z \varphi, \qquad J^{\pm}(z) = \rme^{ \pm \rmi \sqrt{2} \varphi(z)},
\label{jj}
\end{equation}
and the primary fields $\phi_{1/2, s_i}$ ($s_i= \pm 1$) by
\begin{equation}
\phi_{1/2, s_i} (z_i) = \rho_{\frac{1}{2},i} : \rme^{ \rmi s_i \varphi(z_i)/\sqrt{2} } :, \qquad \rho_{\frac{1}{2},i} \equiv
\left\{ \begin{array}{cl}
 1 & i: {\rm even} \\
 \rme^{ \rmi \frac{\pi}{2} ( s_i -1)} & i: {\rm odd}
 \end{array}
 \right.,
 \label{phi1}
\end{equation}
where $: \ldots :$ denotes normal ordering. The two-point correlators of these fields are
\begin{eqnarray}
 \langle J^0(z_1) J^0(z_2) \rangle = \frac{1}{ 2 z_{12}^2}, \qquad
  \langle J^+(z_1) J^-(z_2) \rangle = \frac{1}{ z_{12}^2},\\
 \langle  : \rme^{ \rmi s_1 \varphi(z_1)/\sqrt{2} } :   : \rme^{ \rmi s_2 \varphi(z_2)/\sqrt{2} } : \rangle = \delta_{s_1+s_2,0} \, z_{12}^{ s_1 s_2/2},
\end{eqnarray}
where $z_{12} \equiv z_1 - z_2$ as above. Plugging (\ref{phi1}) into (\ref{psi1}), one gets \cite{cft-book}
\begin{eqnarray}
\psi_{s_1, \dots, s_N} (z_1, \dots, z_N) = \rho_{\frac{1}{2}} \prod_{i < j}^N (z_i - z_j)^{ s_i s_j /2},
\label{psi2}\\
\rho_{\frac{1}{2}} \equiv \rme^{ \rmi \frac{\pi}{2} \sum_{i: {\rm odd}} ( s_i -1) }, \qquad N: {\rm even}, \qquad
\sum_{i=1}^N s_i = 0.\nonumber
\end{eqnarray}
The Marshall sign factor $\rho_{\frac{1}{2}}$ ensures that $\psi$ is a singlet state, as required by (\ref{dec2}).

The Hamiltonian $H_{1, \frac{1}{2}}$, whose ground state is given by (\ref{psi2}), can be constructed using the results of the previous section. For $|z_i|=1$, $\forall i$, we find
\begin{eqnarray}
{\cal C}_{1, \frac{1}{2}}^{i, a} &  = & \frac{2}{3} \sum_{j (\neq i)}^N  w_{i j} \left( t^a_{j}  + \rmi \vep_{a b c} \, t^b_i  t^c_{j} \right), \label{ci} \\
H_{1, \frac{1}{2}}^{(i)}  & = & - \frac{2}{3} \left( \frac{3}{4}
\sum_{ j ( \neq i)}^N w_{i j}^2
+  \sum_{ j_{1} \neq j_2 ( \neq i)}^N w_{i j_1}   \, w_{i j_2} \, t^{a}_{j_1}  \,  t^{a}_{j_2}
+ \sum_{ j  (\neq i)}^N w_{i j}^2   \;
t^{a}_i \, t^a_j   \right),
\label{qqi}  \\
{4\cal H}_{1/2} & \equiv &
H_{1, \frac{1}{2}}  =   - \frac{2}{3} \left[ \frac{3}{4}
\sum_{ i \neq j}^N  w_{i j}^2 + \sum_{ i \neq j}^N \left(
w_{i j}^2
+ \sum_{ k ( \neq i, j )} w_{k i} \, w_{k j } \right)  t^{a}_i \, t^{a}_{j}   \right].
\label{htot22}
\end{eqnarray}

\subsection{Relation to the Haldane-Shastry model} \label{sec:keq1HS}

In 1988, Haldane and Shastry considered the state obtained by the Gutzwiller projection of the one dimensional Fermi state at half filling \cite{H88,S88} (see \cite{ThesisTalstra} for a review)
\begin{equation}
 | \psi_G \rangle = \prod_i ( 1- n_{i, \uparrow} n_{i, \downarrow} ) |{\rm FS} \rangle,
 \label{ss0}
\end{equation}
where $n_{i, s} = c^\dagger_{i, s} c_{i,s} \; (s = \uparrow, \downarrow, i=1, \dots, N) $ are the occupation operators of an electron in a periodic chain with $N$ sites, and $|{\rm FS} \rangle = \prod_{|k| < k_F} \hat{c}^\dagger_k |0 \rangle$ is the Fermi state at half filling. The state (\ref{ss0}) has a unique spin per site because the charge degrees of freedom are frozen by the Gutzwiller projection. These authors then wrote (\ref{ss0}) in terms of hard-core boson variables. Mapping the spin up (down) into the empty (occupied) states of a hard-core boson, they found
\begin{equation}
 | \psi_G \rangle = \sum_{n_1 < \dots < N/2} \psi(n_1, \dots, n_{N/2}) \; b^\dagger_{n_1} \dots b^\dagger_{n_{N/2}} \, |0 \rangle,
 \label{ss0b}
\end{equation}
where $b^\dagger_n$ is the hard-core boson creation operator at the $n^{\rm th}$ site and
\begin{eqnarray}
\psi(n_1, \dots, n_{N/2}) \propto   \rme^{ \rmi \pi  \sum_i n_i  }  \; \prod_{n_i > n_j}^{N/2}  \left[
\sin \frac{ \pi (n_i-n_j)}{N} \right]^{2}.
\label{wave3}
\end{eqnarray}
Using this equation, Haldane and Shastry showed that $|\psi_G\rangle$ is the ground state of the spin Hamiltonian
\begin{equation}
{\cal H}_{\rm HS} = - \sum_{i \neq j}  \frac{z_i z_j }{z_{ij}^2 } \,
(\mathcal{P}_{ij}-1),
\label{hs1}
\end{equation}
where $\mathcal{P}_{ij}$ is the permutation operator of the spins at the positions $i$ and $j$. The permutation can be expressed in terms of spin operators as $\mathcal{P}_{ij} = 2 t^a_i t^a_j +1/2$, so (\ref{hs1}) takes the form
\begin{equation}
{\cal H}_{\rm HS} = -  \sum_{i \neq j}  \frac{z_i z_j }{z_{ij}^2 } \, \left(2  t^a_i t^a_j - \frac{1}{2} \right).
\label{hs2}
\end{equation}
In \cite{CS10}, it was shown that the wave function (\ref{wave3}) corresponds to the state (\ref{psi2}) for the choice $z_n = \rme^{ 2 \pi \rmi n/N}$. Indeed, let us denote by $q_i =0,1$ the occupation number of a hard-core boson at the site $i$. The mapping between the spin variable $s_i = \pm 1$ and $q_i$ is given by $s_i = 1 - 2 q_i$ and one can easily show that (\ref{wave3}) becomes
\begin{eqnarray}
\psi(s_1, \dots, s_N) \propto \rme^{ \rmi \frac{\pi}{2} \sum_{i: {\rm odd} } s_i } \prod_{n > m}^N \left[
\sin \frac{ \pi (n-m)}{N} \right]^{ s_n s_m/2}, \qquad \sum_{i=1}^N s_i = 0,
\label{wave2}
\end{eqnarray}
where the constraint $\sum_{i=1}^N s_i = 0$ guarantees that the total third component of the spin vanishes. This expression is proportional to (\ref{psi2}).

From these considerations one must expect the HS Hamiltonian (\ref{hs1}) to be intimately related to the Hamiltonian (\ref{htot22}). To show this relation in detail, we need some simple identities. For generic values of $z_i$, one can show that
\begin{equation}
w_{ij}^2  = 1 + 4 \frac{z_i z_j}{z_{ij}^2} \label{wid1}
\end{equation}
and
\begin{equation}
w_{k i } w_{k j} + w_{i j } w_{i k} + w_{j k} w_{j i} = 1, \qquad i \neq j \neq k.
\label{wid}
\end{equation}
The latter identity implies
\begin{eqnarray}
\sum_{ k ( \neq i, j )}^N  w_{k i} \, w_{k j } & = & 2 w_{ij}^2 + w_{ij} ( c_i - c_j) + N-2, \qquad
c_i \equiv \sum_{k (\neq i)}^N w_{ki},
\label{wid2}
\end{eqnarray}
which plugged into (\ref{htot22}) yields
\begin{eqnarray}
\fl{\cal H}_{1/2} & = & - \frac{1}{2} \sum_{ i \neq j} \frac{ z_i z_j}{ z^2_{ij} } + \frac{N}{4}
 - \frac{(N+1)}{6} T^a T^a - 2  \sum_{i \neq j} \left[ \frac{z_i z_j }{z_{ij}^2 } + \frac{ w_{ij} ( c_i - c_j)}{ 12} \right]
 t^a_i \, t^a_j,
 \label{hbis}
\end{eqnarray}
where $T^a$ is the total spin operator defined in (\ref{dec2}). In the uniform case, we have
\begin{equation}
c_i = 0 \;\; \forall i, \qquad
\sum_{ i \neq j} \frac{ z_i z_j}{ z_{ij} ^2} = - \frac{ N ( N^2 -1)}{ 12},
\label{uniform}
\end{equation}
so
\begin{equation}
 {\cal H}_{1/2} =  {\cal H}_{\rm HS} - E_{0}  - \frac{(N+1)}{6} T^a T^a , \qquad E_{0} = - \frac{N^3 + 2 N}{12},
 \label{HS1-2}
\end{equation}
where $E_{0}$ is the ground state energy of ${\cal H}_{\rm HS}$. The positivity of ${\cal H}_{1/2}$ implies a lower bound for
the energies of the eigenstates of ${\cal H}_{\rm HS}$ with total spin $S$, i.e.,
\begin{equation}
E_{S} - E_0 \geq \frac{1}{6} (N+1) S ( S+1),
\end{equation}
which is indeed satisfied by the lowest eigenenergy $E_S$ in the total spin sector $S \leq N/2$,
\begin{equation}
E_S - E_0 = \frac{1}{6} S \, ( 2 + 3 N S - 2 S^2).
\end{equation}
The latter formula can be derived from the general expression for the eigenvalues of (\ref{hs1}). The eigenvalues were found in \cite{H91} and can be expressed in terms of a set $ \{ m_j \}$ of nonconsecutive integers $m_j \in \{1,2, \dots, N-1\}$, $m_{j+1} \geq m_j +1$, called rapidities \cite{yang}
\begin{eqnarray}
E( \{ m_j \}) = \sum_j m_j ( m_j - N).
\end{eqnarray}
Thus, for example, the ground state corresponds to the set $\{ m_j \} = (1,3,5, \dots, N-1)$ ($N$ even), and the
lowest excited state of total spin 1 to $\{ m_j \} = (2,4,6, \dots, N-2)$.

In the nonuniform case, where $c_i \neq 0$ for some $i$, we get a generalization $\tilde{{\cal H}}_{\rm HS}$ of the HS model. Specifically,
\begin{equation}
{\cal H}_{1/2} =  \tilde{\cal H}_{\rm HS} - \tilde{E}_{0}
- \frac{(N+1)}{6} T^a T^a ,
\label{HS1-2tilde}
\end{equation}
where
\begin{equation}
\tilde{{\cal H}}_{\rm HS} = - \sum_{i \neq j}  \left[ \frac{z_i z_j }{z_{ij}^2 } + \frac{ w_{ij} ( c_i - c_j)}{ 12} \right] \, ( \mathcal{P}_{ij} -1)
\label{hs3}
\end{equation}
and the ground state energy is
\begin{equation}
\tilde{E}_0 = \sum_{i \neq j} \left[ \frac{z_i z_j }{z_{ij}^2 } + \frac{ w_{ij} (c_i - c_j)}{ 24} \right] - \frac{N}{4}.
\end{equation}
Note that the coupling strength in (\ref{hs3}) is more complicated than just the inverse of the square of the distance between the spins measured along the chord. In \cite{CS10}, a different approach based on the KZ equation also led to the Hamiltonian $\tilde{\mathcal{H}}_{\rm HS}$, and it was demonstrated that $\psi$ is an eigenstate of $\tilde{\mathcal{H}}_{\rm HS}$ with eigenvalue $\tilde{E}_0$. We provide the details of this derivation in \ref{sec:alternative}.

The spectrum of the uniform HS model exhibits an enhanced symmetry given by a Yangian Hopf algebra \cite{D85} generated by the total spin operator $T^a$ and the operator \cite{yang}
\begin{equation}
\Lambda^a = \frac{\rmi}{2} \sum_{i \neq j} w_{ij}  \, \vep_{a b c} \, t^b_i \, t^c_j.
\label{yan}
\end{equation}
The operators $T^a$ and $\Lambda^a$ all commute with the Hamiltonian (\ref{hs1}) but they do not commute among themselves, and they are the key to compute the complete spectrum mentioned above. Observe that $\Lambda$ can be constructed from the operator (\ref{ci}),
\begin{equation}
\Lambda^a = \frac{3}{4} \sum_{i}  {\cal C}_{1, \frac{1}{2}}^{i, a}.
\label{newy}
\end{equation}
It follows that $\psi$ is an eigenstate of (\ref{newy}) (with eigenvalue zero) in both the uniform and the nonuniform cases. In the nonuniform case, however, the Yangian symmetry is broken, meaning that (\ref{newy}) does not commute with (\ref{hs3}). In the last part of \ref{sec:alternative}, we show that $\psi$ is also an eigenstate (with eigenvalue zero) of the operator
\begin{equation}\label{H3}
\fl H_3 = - \rmi \sum_{i\neq j\neq k} \frac{z_i z_j z_k}{ z_{ij} z_{jk} z_{ki}} \; \vep_{abc}t_i^at_j^bt_k^c +
\sum_{i\neq j} t_i^at_j^a \left( - \frac{1}{2} \sum_{k (\neq i)} w_{ki} + \frac{17}{8} w_{ij} \sum_{k (\neq i)} w_{ki}^2 \right),
\end{equation}
which is a generalization of the Inozemtsev invariant \cite{I90} to the nonuniform case. In the uniform case, this operator commutes with the Hamiltonian (\ref{hs1}) \cite{I90}, but $H_3$ does not commute with $\tilde{\cal H}_{\rm HS}$ in general.

To illustrate the breaking of symmetries, we compute the spectrum for the three different choices of $z_i$ considered in \cite{CS10}: uniform, dimer and random. The dimer model is defined by the following choice of parameters
\begin{equation}
z_j = \rme^{ 2 \pi \rmi [ j + (-1)^j \delta]/N}, \qquad 0 \leq \delta \leq \frac{1}{2}, \qquad j =1, 2 \dots, N,
\label{dim}
\end{equation}
where $\delta$ characterizes the dimerization of the ground state. The value of $c_j$ defined in (\ref{wid2}) is
\begin{equation}
c_j = \rmi \frac{N}{2} (-1)^{j+1} \, \tan( \pi \delta),
\end{equation}
which only vanishes in the uniform case $\delta=0$. The random model is defined by the choice
\begin{equation}
z_j = \rme^{ 2 \pi \rmi ( j + \phi_j)/N}, \qquad - \frac{\delta}{2} \leq \phi_j <  \frac{\delta}{2}, \qquad j =1, 2 \dots, N,
\label{ran}
\end{equation}
where $\phi_j$ is a random variable uniformly distributed in the interval $[ - \frac{\delta}{2}, \frac{\delta}{2}]$. The parameter $\delta$ determines the degree of disorder in the ensemble. The choice (\ref{ran}) gives the random HS version of the spin 1/2 antiferromagnetic Heisenberg model (AFH) with random exchange couplings. The latter model belongs to the universality class known as the random singlet phase. The Renyi entropies of this phase differ from that of the uniform AFH model given by the $SU(2)_1$ model \cite{RM04,RM09}.

\begin{table}
\caption{\label{compare}Spectrum of the generalized HS Hamiltonian (\ref{hs3}) for $N=6$ sites and three different choices of $z_j$: Left: uniform (\ref{uniform}). Middle: dimer with $\delta=0.1$ (\ref{dim}). Right: random for one particular realization of $\phi_j$ and $\delta=1$ (\ref{ran}). $E$ is the energy, $S$ is the spin, $p_u$ and $p_d$ are momenta (see text), and $\#=2S+1$ is the number of (degenerate) states within each of the spin multiplets.}
\begin{indented}
\lineup
\item[]\begin{tabular}{@{}llll}
\br
$E$ & $S$ & $\frac{3p_u}{\pi}$ & \# \\
\mr
   -19 &    0 &    \ 3& 1 \\
   -16 &    0 &    \ 0& 1 \\
       &    1 &    \ 0& 3 \\
   -14 &    1 &     -2& 3 \\
       &    1 &    \ 2& 3 \\
   -13 &    0 &     -1& 1 \\
       &    0 &    \ 1& 1 \\
       &    1 &     -1& 3 \\
       &    1 &    \ 1& 3 \\
   -10 &    1 &    \ 0& 3 \\
    \0-9 &    0 &  \ 3& 1 \\
       &    1 &    \ 3& 3 \\
       &    2 &    \ 3& 5 \\
    \0-8 &    1 &   -2& 3 \\
       &    1 &    \ 2& 3 \\
       &    2 &     -2& 5 \\
       &    2 &    \ 2& 5 \\
    \0-5 &    2 &   -1& 5 \\
       &    2 &    \ 1& 5 \\
    \0\ 0 &    3 &   \ 0& 7 \\
\br
\end{tabular}
\hfill
\begin{tabular}{@{}llll}
\br
$E$ & $S$ & $\frac{3p_d}{\pi}$ & \# \\
\mr
  -19.9502&    0&    \ 0&  1\\
  -16.6431&    1&    \ 0&  3\\
  -16.3598&    0&    \ 0&  1\\
  -14.6851&    1&     -2&  3\\
  &    1&    \ 2&  3\\
  -13.3167&    0&   -2&  1\\
  &    0&    \ 2&  1\\
  -13.2717&    1&   -2&  3\\
  &    1&    \ 2&  3\\
  -10.3401&    1&    \ 0&  3\\
   \0-9.3167&    2&    \ 0&  5\\
   \0-8.9669&    1&    \ 0&  3\\
   \0-8.9570&    0&    \ 0&  1\\
   \0-8.3167&    2&   -2&  5\\
   &    2&    \ 2&  5\\
   \0-7.9934&    1&   -2&  3\\
   &    1&    \ 2&  3\\
   \0-5.0000&    2&   -2&  5\\
   &    2&    \ 2&  5\\
   \0-0.0000&    3&   \ 0&  7\\
\br
\end{tabular}
\hfill
\begin{tabular}{@{}lll}
\br
$E$ & $S$ & \# \\
\mr
  -22.9229&    0&  1\\
  -19.8148&    1&  3\\
  -18.1316&    0&  1\\
  -16.8832&    1&  3\\
  -15.8047&    1&  3\\
  -15.2108&    0&  1\\
  -14.9450&    1&  3\\
  -14.0555&    1&  3\\
  -12.7601&    0&  1\\
  -11.4068&    2&  5\\
  -10.4704&    1&  3\\
  -10.0944&    2&  5\\
 \0-9.0416&    1&  3\\
 \0-8.8204&    0&  1\\
 \0-8.0155&    1&  3\\
 \0-7.7381&    1&  3\\
 \0-7.5353&    2&  5\\
 \0-5.0000&    2&  5\\
 \0-4.8865&    2&  5\\
 \0-0.0000&    3&  7\\
\br
\end{tabular}
\end{indented}
\end{table}

The spectrum of the generalized HS Hamiltonian (\ref{hs3}) for these three models is given in table \ref{compare} for $N=6$. The Hamiltonian $\tilde{{\cal H}}_{\rm HS}$, the dot product $T^aT^a$ of the total spin operator with itself and the $z$-component $T^0$ of the total spin operator all commute with each other. It is hence possible to diagonalize these operators simultaneously and label the resulting states by their energy $E$, spin $S$ and z-component of the total spin $S_z$. We have omitted $S_z$ from the table, because $T^\pm$ also commute with the Hamiltonian, and it follows that the $2S+1$ states within each spin multiplet are degenerate. Instead we write only the number of states in each multiplet. In the uniform case, the momentum operator also commutes with $T^a$ and the Hamiltonian due to invariance under translation by one site. The momentum operator can be defined as $p_u=-\rmi\ln(\mathcal{T})$, where
\begin{equation}\label{trans}
\mathcal{T}=\mathcal{P}_{N,N-1}\mathcal{P}_{N-1,N-2}\ldots \mathcal{P}_{2,1}
\end{equation}
is the translation operator and $\mathcal{P}_{i,j}$ is the operator that permutes spins $i$ and $j$ as defined just after (\ref{hs1}). The momentum eigenvalues in units of $\pi/3$ are given in the table. The fact that the chain is invariant under the flip operation
\begin{equation}\label{flip}
\mathcal{P}_{N,1}\mathcal{P}_{N-1,2}\ldots\mathcal{P}_{N/2+1,N/2}
\end{equation}
explains why states with momentum eigenvalue $\pm\pi/3$ or $\pm2\pi/3$ always appear in degenerate pairs. In the dimer case, $p_u$ does not commute with the Hamiltonian, but since the chain is invariant under translation by two sites, we can instead add the momentum operator $p_d=-\rmi\ln(\mathcal{T}^2)$ to the set of mutually commuting operators. Also in this case, states with $p_d$ eigenvalue $\pm2\pi/3$ always appear in degenerate pairs due to invariance under the flip operation (\ref{flip}). For the random case, there is no translational symmetry. Looking at the degeneracies between different spin multiplets, we observe that the number of degeneracies decreases in going from the uniform to the dimer case and in going from the dimer to the random case. In the dimer case, the Yangian symmetry is broken and only degeneracies between pairs with opposite $p_d$ eigenvalues occur. In the random case, there are even less symmetries, and all spin multiplets have different energies.

This subsection shows that the generalization of the uniform HS model embraces several interesting models within a common framework. We shall show below that more information about this model can be obtained exploiting the algebraic decoupling equations.

\subsection{Spin correlators}\label{sec:keq1CF}

In 1987, Gebhard and Vollhardt computed the two-point spin correlator in the Gutzwiller state (\ref{ss0}) for an infinite chain, obtaining the exact formula \cite{GV87}
\begin{equation}
\langle t^a_n \, t^b_0 \rangle = (-1)^n \delta_{a b} \frac{ {\rm Si}(  \pi n)}{ 4 \pi n}, \qquad t^a_n = \frac{1}{2} c^\dagger_{n, s} \, \sigma^a_{s s'} c_{n, s'},
\label{ss1}
\end{equation}
where ${\rm Si}(z)$ is the sine integral function
\begin{equation}
{\rm Si}(z) = \int_0^z \rmd t \, \frac{ \sin(t)}{t}.
\label{Si}
\end{equation}
The appearance of the term $\delta_{ ab}$ reflects the $SU(2)$ symmetry of the Gutzwiller state.
It may be worthwhile to notice that this result triggered the Haldane-Shastry works \cite{H88,S88}. Using the asymptotic expression of the sine integral function
\begin{equation}
 {\rm Si}(\pi n ) \simeq  \frac{\pi}{2} + \frac{ (-1)^{n+1}}{ \pi n},  \qquad n \in \mathbb{Z}, \qquad n \gg 1,
\label{Si2}
\end{equation}
one obtains
\begin{equation}
\langle t^a_n \, t^b_0 \rangle \simeq \delta_{a b} \left[ \frac{(-1)^n}{ 8 n} - \frac{1}{ 4 \pi^2 n^2} \right], \qquad n \gg 1.
\label{Si3}
\end{equation}

One can also derive (\ref{Si3}) from the underlying CFT
that describes the long distance behaviour of the HS model that
is given by the $SU(2)_1$ WZW model. The spin operators $t^a_n$
can be expressed in terms of the current operators $J^a(z)$ (\ref{jj}) and their antiholomorphic version $\bar{J}^a(\bar{z})$
as \cite{AH87,A88}
\begin{equation}
\eqalign{t^0_n \simeq J^0(z_n) + \bar{J}^0(\bar{z}_n) + (-1)^n \sqrt{ \frac{\pi}{2}} \cos \left( \frac{ \varphi(z_n, \bar{z}_n) }{\sqrt{2}} \right),
\label{Si4} \\
t^\pm_n \simeq J^\pm(z_n) + \bar{J}^\pm(\bar{z}_n) + (-1)^n \sqrt{ \frac{\pi}{2}} \exp \left( \pm \rmi \frac{ \varphi(z_n, \bar{z}_n) }{\sqrt{2}} \right),}
\end{equation}
where $z_n = 2 \pi \rmi n$ and $\varphi(z, \bar{z}) = \varphi(z) + \bar{\varphi}(\bar{z})$.
It is interesting to compare the spin correlator of the HS model with that of the AFH model. The latter correlator behaves as \cite{AG89,E96}
\begin{equation}
\langle t^a_n \, t^b_0 \rangle = \delta_{a b} \left[ C \frac{ (-1)^n(\ln(n))^{1/2}}{n} - \frac{1}{ 4 \pi^2 n^2} \right]
\label{corafh}
\end{equation}
where $C$ is a nonuniversal constant. The logarithmic correction to the leading $1/n$ term is due to the marginal irrelevant operator $J^a \bar{J}^a$, which added to the action of the WZW model gives the long distance behaviour of the AFH model \cite{AH87}. In this respect, the HS model is at the fixed point of the renormalization group, so that the continuum Hamiltonian does not contain the marginal operator $J^a \bar{J} ^a$, which explains why the logarithmic corrections in the correlator (\ref{Si3}) are absent.

\subsubsection{Algebraic equations for spin correlators}\label{sec:eqspincor}

We shall next derive (\ref{ss1}) and several other expressions for spin correlators from (\ref{ss20a}). From (\ref{ci}), we find
\begin{equation}
A_{1,\frac{1}{2}}^{i, a}
= \frac{2}{3} \sum_{k (\neq i)} w_{ik} t^a_k,
\qquad
Q_{1,\frac{1}{2}}^{i, a}
= \frac{2}{3} \sum_{k (\neq i)} w_{ik} \, \rmi \, \vep_{abc} t^b_i t^c_k,
\end{equation}
where it is assumed that $|z_i|=1$, such that $w_{ij}$ is imaginary. Note that $A_{1,\frac{1}{2}}^{i, a}$ and $Q_{1,\frac{1}{2}}^{i, a}$ are one-body and two-body operators, respectively. For the spin $1/2$ case, a product of spin operators acting on the same spin can always be written as a sum of terms containing at most one spin operator each, so it is sufficient to consider spin correlators for which all the spin operators act on different spins. We hence assume that all the indices $i_1 \ldots i_n$ in $T_{i_1\ldots i_n}^{a_1\ldots a_n}$ are different and also different from the index $i$. $T_{i_1\ldots i_n}^{a_1\ldots a_n}$ is then an $n$-body operator. If we choose $n$ odd in (\ref{ss20a}), we get $3^{n+1}N(N-1)\cdots (N-n)$ linear equations involving only ($n+1$)-point and ($n-1$)-point spin correlators. Starting from $n=1$, this allows us to successively compute spin correlators with higher and higher numbers of operators as we shall see below.

Finally, we mention the important result
\begin{eqnarray}
w_{ij} \langle t^a_i t^a_j \rangle + \sum_{k ( \neq i,j ) } w_{ik} \langle t^a_j t^a_k \rangle +  \frac{3}{4} w_{ij} & = & 0, \qquad i \neq j,
\label{ss4}
\end{eqnarray}
which follows from (\ref{ss20a}) with $n=1$ and $b_1=a_1$. (\ref{ss4}) is a system of linear equations for the two-point correlators $\langle t^a_i t^a_j \rangle$ with $i\neq j$ that can be solved numerically for generic values of the parameters $z_j$, and analytically in special cases. In the following paragraphs, we shall investigate some particular solutions.

\subsubsection{Two-point spin correlators for the infinite chain}

Let us choose
\begin{equation}
z_j = \rme^{ 2 \pi \rmi x_j/N}, \qquad x_j \in \mathbb{R}, \qquad j = - \frac{N}{2} + 1, \dots, \frac{N}{2}.
\end{equation}
Then, in the limit
\begin{equation}
|x_i| \ll N,  \, \forall i, \Longrightarrow w_{i j} = - \rmi \cot \frac{ \pi (x_i- x_j)}{N} \rightarrow - \frac{\rmi N}{\pi} \frac{ 1}{x_i - x_j},
\end{equation}
(\ref{ss4}) becomes
\begin{equation}
C_{ij} + \sum_{k (\neq i,j) } \frac{ x_{ij}}{ x_{ik}} C_{jk} + \frac{3}{4} =0, \qquad C_{ij} = \langle t^a_i t^a_j \rangle, \qquad i \neq j ,
\label{ss5}
\end{equation}
where the equality is reached in the limit $N \rightarrow \infty$. In the uniform case, i.e.\ $x_n = n \; (n \in \mathbb{Z})$, translational invariance implies that $C_{ij}$ only depends on the difference $|i-j|$, hence it is sufficient to consider the case $i=0$ and $j=n$ in (\ref{ss5}),
\begin{equation}
C(n) + \sum_{k (\neq 0) } \frac{ n}{k} \, C(n-k) =0, \qquad n \neq 0, \qquad C(n) = C_{0,n}= C_{n,0},
\label{ss6}
\end{equation}
where we have used that $C(0) = \langle t^a_0 t^a_0 \rangle = 3/4$.
To solve these equations, we introduce the Fourier transform of $C(n)$,
\begin{equation}
G(q) = \sum_{n=- \infty}^\infty \rme^{ \rmi n q} \, C(n) , \qquad C(n) = \int_{- \pi}^\pi \frac{ \rmd q }{ 2 \pi} \rme^{- \rmi n q} G(q).
\label{ss7}
\end{equation}
Using the identities
\begin{equation}
\sum_{n = - \infty}^\infty n \, \rme^{ \rmi n q} = - 2 \pi \rmi \, \frac{\rmd}{\rmd q}\delta(q), \qquad
\sum_{n = - \infty }^{\infty}  \frac{1}{n} \, \rme^{ \rmi n q} = \rmi \pi \left( \sign(q) - \frac{ q}{\pi} \right),
\label{ss8}
\end{equation}
(\ref{ss6}) turns into the differential equation
\begin{equation}
\pi \left( \sign(q) - \frac{ q}{\pi} \right)  \frac{ \rmd G(q)}{\rmd q} = \frac{3}{4}.
\label{ss9}
\end{equation}
The $SU(2)$ invariance of the ground state implies that $G(0) = \langle T^a \, t^a_0 \rangle = 0$, hence integrating (\ref{ss9})
yields
\begin{equation}
 G(q) = - \frac{3}{4} \ln \left( 1 - \frac{ |q|}{\pi} \right).
\label{ss10}
\end{equation}
Finally, $C(n)$ is given by the inverse Fourier transform,
\begin{eqnarray}
C(n) & = - \frac{3}{4} \int_{- \pi}^\pi \frac{\rmd q}{2 \pi} \rme^{- \rmi n q} \ln \left( 1 - \frac{ |q|}{\pi} \right)\\
& = - \frac{3}{4 \pi } \int_{0}^\pi \rmd q \,  \cos( n q)  \ln \left( 1 - \frac{ q}{\pi} \right)\nonumber\\
& = \frac{3}{4 \pi^2 n } \int_{0}^\pi \rmd q \, \sin( n q)  \frac{1}{ 1 - q/\pi} = \frac{3}{4 \pi n } \cos( \pi n) \int_{0}^{ n \pi} \rmd t  \, \frac{ \sin( t)}{t}.
\nonumber
\end{eqnarray}
Since $\langle t_n^at_0^b\rangle=\delta_{ab}\langle t_n^ct_0^c\rangle/3$ due to the $SU(2)$ symmetry, (\ref{ss1}) follows.

\subsubsection{Two-point spin correlators for finite chains}

Let us next show that (\ref{ss4}) can be solved analytically for a uniform spin chain with a finite number of sites $N$, and not
only in the large $N$ limit, as done above. We write
\begin{equation}
z_j = z^j   \; \; (j =1, \dots, N), \qquad z \equiv \rme^{ 2 \pi \rmi/N},
\end{equation}
so that (\ref{ss4}) becomes
\begin{equation}
\frac{ 1 + z^{-n}}{ 1 - z^{-n}} C(n) + \sum_{k =1}^{N-1} \frac{ 1 + z^{-k}}{ 1 - z^{-k}} C(n-k) = 0, \qquad n \neq N.
\end{equation}
If $n= N/2$, the first term of this equation vanishes, which leads us to split this equation as
\begin{eqnarray}
C(n)  +  \sum_{k =1}^{N-1} \frac{ 1 - z^{-n}}{ 1 + z^{-n}}  \frac{ 1 + z^{-k}}{ 1 - z^{-k}} C(n-k) = 0, \qquad n \neq \frac{N}{2}, N,
\label{ss11a} \\
  \sum_{k =1}^{N-1} \frac{ 1 + z^{-k}}{ 1 - z^{-k}} C(N/2 -k) = 0.
\label{ss11b}
\end{eqnarray}
To solve these equations, one introduces the finite Fourier transform (recall (\ref{ss7}))
\begin{equation}
G_q = \sum_{n=1}^N z^{ n q} \, C(n) , \qquad C(n) = \frac{1}{N} \sum_{q=1}^N \ z^{ - n q} G_q,
\label{ss12}
\end{equation}
where $G_N = \sum_n C(n) =0$. Now we multiply (\ref{ss11a}) by $z^{n q}$ and sum over $n \neq N/2, N$ to obtain
\begin{equation}
\fl G_q - \frac{3}{4} - (-1)^q C(N/2) + \frac{1}{N} \sum_{q' =1}^N G_{q'} \sum_{n = 1 (\neq N/2)}^N z^{ n ( q - q' )}
 \frac{ 1 - z^{-n}}{ 1 + z^{-n}} \sum_{ k=1}^{N-1} z^{ k q' } \frac{ 1 + z^{-k}}{ 1 - z^{-k}} = 0,
\label{ss13}
\end{equation}
while (\ref{ss11b}) becomes
\begin{equation}
\sum_{q' =1}^N G_{q'} \; (-1)^{q'} \sum_{ k=1}^{N-1} z^{ k q' } \frac{ 1 + z^{-k}}{ 1 - z^{-k}} = 0.
\label{ss14}
\end{equation}
The two sums appearing in (\ref{ss13}) are the discrete version of the integrals (\ref{ss8}), and they are given by
\begin{eqnarray}
\fl A(q) \equiv \frac{1}{N} \sum_{ k=1}^{N-1} z^{ k q } \,  \frac{ 1 + z^{-k}}{ 1 - z^{-k}} = 1 + \delta_{q, N} - \frac{2 q}{N}, \qquad 1 \leq q \leq N,
\label{ss15a} \\
\fl B(q) \equiv \frac{1}{N} \sum_{ n=1 (\neq N/2)}^{N} z^{ n q } \frac{ 1 - z^{-n}}{ 1 + z^{-n}} = (-1) ^q \left( \sign (q) - \frac{2 q}{N} \right), \qquad |q| \leq N-1,
\label{ss15b}
\end{eqnarray}
in terms of which (\ref{ss13}) and (\ref{ss14}) become
\begin{eqnarray}
G_q - \frac{3}{4} + (-1)^q \, a + (-1)^q N \sum_{q'=1}^N G_{q' } A_{q' } (-1)^{q' } \; \sign ( q- q' ) = 0, \label{ss16a} \\
 \sum_{q'=1}^N G_{q' } A_{q' } (-1)^{q' } = 0, \label{ss16b}
\end{eqnarray}
where $a$ is a constant given by
\begin{equation}
a = 2 \sum_{q=1}^N G_{q } A_{q } \,  q (-1)^{q } - C(N/2).
\label{ss17}
\end{equation}
It is convenient to define the quantity
\begin{equation}
g_q = (-1)^q \; G_q \; (q=1, \dots, N),
\end{equation}
which must satisfy
\begin{eqnarray}
g_q - \frac{3}{4} (-1)^q + a + N \sum_{q'=1}^N g_{q' } A_{q' }\; \sign ( q- q' ) = 0, \label{ss17a} \\
 \sum_{q'=1}^N g_{q' } A_{q' } = 0. \label{ss17b}
\end{eqnarray}
Taking the difference for two consecutive values $q$ and $q+1$ in (\ref{ss17a}) yields
\begin{equation}
g_{q+1} - g_q + N( g_q A_q + g_{q+1} A_{q+1}) + \frac{3}{2} (-1)^q = 0, \qquad 1 \leq q \leq N-1,
\end{equation}
which, when translated into the $G_q$ quantities and using (\ref{ss15a}) for $A_q$, becomes
\begin{equation}
\left( 1 - \frac{ 2 q +1}{N} \right) ( G_{q+1} - G_q ) = \frac{3}{2N},
\qquad 1 \leq q \leq N-1,
\end{equation}
and hence
\begin{equation}
G_q = - \frac{3}{2} \sum_{q' = q}^{N-1} \frac{1}{ N- 1 - 2 q' }, \qquad q= 1, \dots, N-1, \qquad G_N = 0.
\label{ss18}
\end{equation}
One can check that (\ref{ss17}) is also satisfied and that the parameter $a$ is equal to $3/4$. Plugging this expression into (\ref{ss12}), one finally finds the correlator
\begin{equation}
\langle t^a_n \, t^b_0 \rangle = \frac{1}{3} \delta_{a b} \; C(n) =  \delta_{ ab} \frac{ (-1)^n}{4N\sin\left( \pi n/N \right)}
\sum_{m=1}^{N/2}  \frac{\sin\left[2\pi n(m - \frac{1}{2})/N \right]}
{ m - \frac{1}{2} }.
\label{ss19}
\end{equation}
In the limit $N \rightarrow\infty$, one recovers (\ref{ss1}). Moreover, one can take the limit $N \gg 1$ and $n \gg 1$ with the ratio $n/N$ kept constant, obtaining
\begin{equation}
\langle t^a_n \, t^b_0 \rangle \simeq \delta_{ab}
\left[ \frac{ \pi (-1)^n}{ 8 N \sin \left(\pi n/N \right) }
- \frac{ 1}{ 4 N^2 \sin^2 \left(\pi n/N \right) } \right].
\label{ss20}
\end{equation}

\begin{figure}
\begin{indented}
\item[]\includegraphics[width=0.6\textwidth]{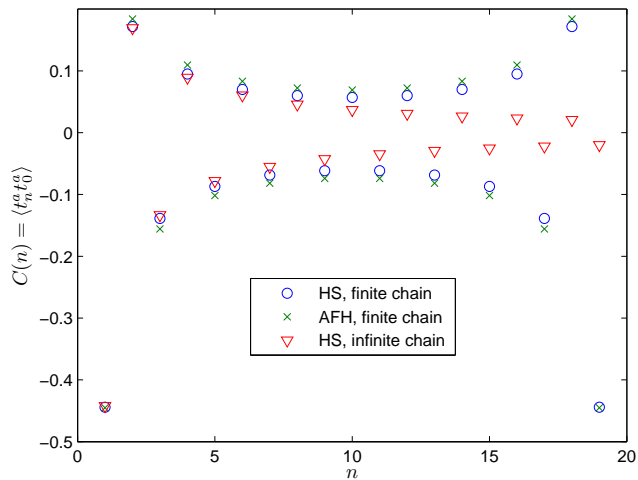}
\end{indented}
\caption{Two-point spin correlation function $C(n) = \langle t^a_n t^a_0 \rangle \;(n=1, \dots, N-1)$ for the uniform HS wave function for a chain with $N=20$ sites. The results correspond to (i) the exact formula (\ref{ss19}), (ii) the antiferromagnetic Heisenberg model with nearest neighbour couplings, and (iii) the thermodynamic limit (\ref{ss1}).}
\label{comparison}
\end{figure}

Equation (\ref{ss20}) can be obtained from the bosonization formulas (\ref{Si4}) with the operators
$J^a(z)$, $\bar{J}^a(\bar{z})$ and $\cos(\varphi(z, \bar{z})/\sqrt{2})$ being defined on a cylinder of circumference $2\pi N$, i.e., $z = z + 2\pi \rmi N$. The two-point correlators needed to reproduce (\ref{ss20})
can be deduced using the conformal transformation from this cylinder to the complex plane, $z \rightarrow w = \rme^{ z/N}$ , which yields
\begin{eqnarray*}
\fl\langle J^0(2 \pi \rmi n_1) J^0(2 \pi \rmi n_2) \rangle_{\rm cyl} = - \left[8 N^2 \sin^2 \left( \frac{ \pi (n_1 - n_2)}{N} \right) \right]^{-1},\\
\fl\left\langle \cos\left(\frac{ \varphi (2 \pi \rmi n_1, - 2 \pi \rmi n_1)}{\sqrt{2}}\right) \cos\left(\frac{ \varphi (2 \pi \rmi n_2, - 2 \pi \rmi n_2)}{\sqrt{2}}\right)\right\rangle_{\rm cyl} =
 \left[ 4 N \left|\sin \left( \frac{ \pi (n_1 - n_2)}{N} \right)\right| \right]^{-1}.
\end{eqnarray*}
As a summary of these results, we present a comparison between the correlators computed with the formulas (\ref{ss1}) and (\ref{ss19}) for the HS model and the correlation function for the AFH model in figure \ref{comparison}.

\subsubsection{Two-point spin correlators in dimerized chains}
\label{sec:dimer}

\begin{figure}
\begin{indented}
\item[]\includegraphics[width=0.6\textwidth]{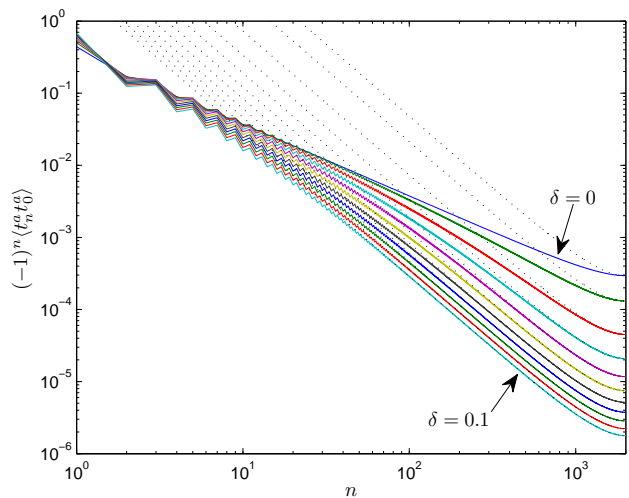}
\end{indented}
\caption{Absolute value of the two-point spin correlation function (i.e.,  $(-1)^n\langle t_n^at_0^a\rangle$) for the dimer model with (starting from above) $\delta=0$, $0.01$, $0.02$, $\ldots$, $0.10$ and $N=4000$ (only the region $1\leq n\leq N/2$ is shown). The dotted lines are all proportional to $[\sin(\pi n/N)]^{-2}$}\label{DimerCor}
\end{figure}

In the dimer case (\ref{dim}), translational invariance by one site is broken, but there is still translational invariance for translation by an even number of sites. From the symmetries of the wave function, it follows that $\langle t_n^at_m^a\rangle=\langle t_{m-n}^at_0^a\rangle$ for $m$ odd and $\langle t_n^at_m^a\rangle=\langle t_{n-m}^at_0^a\rangle$ for $m$ even, so it is sufficient to compute $\langle t_n^a \, t_{0}^a \rangle$ for $n=1,2,\ldots,N$. Figure \ref{DimerCor} provides numerical results for this correlator for a chain with $N=4000$ sites obtained from (\ref{ss4}). For $\delta=0$, we observe that the absolute value of the correlator decays as $n^{-1}$ in the region $1 \ll n \ll N$ as it should according to (\ref{Si3}). For $\delta=0.1$, we find that the absolute value of the correlator decays as $[\sin(\pi n/N)]^{-2}$, when the spins are sufficiently far apart. As $\delta$ decreases towards zero, $n/N$ needs to be closer and closer to one half for this approximation to hold. It is expected that the decay is algebraic for $1 \ll n \ll N$ rather than exponential even if the model is not critical because there are long-range interactions in the Hamiltonian \cite{dpc05,scw05}.

\subsubsection{Four-point spin correlators}

We shall now use (\ref{ss20a}) to compute the four-point spin correlator $\langle t^{a_1}_{i_1} \, t^{a_2}_{i_2} \, t^{a_3}_{i_3} \, t^{a_4}_{i_4} \rangle$. As explained above, we can assume the indices $i_1,\ldots,i_4$ to be different and consider the quantity
\begin{equation}
C^{a_1 a_2 a_3 a_4}_{i_1 i_2 i_3 i_4} \equiv
\left\{
\begin{array}{ll}
\langle t^{a_1}_{i_1} \, t^{a_2}_{i_2} \, t^{a_3}_{i_3} \, t^{a_4}_{i_4} \rangle, & {\rm if} \; \; i_\alpha \neq i_\beta \; \;  \textrm{for all } \alpha \neq \beta  \\
0,  & {\rm if} \; \; i_\alpha = i_\beta \; \;  \textrm{for some }\alpha \neq \beta
\end{array}
\right..
\label{ss21}
\end{equation}
Choosing $n=3$ and $T_3 = t^{a_1}_{i_1} t^{a_2}_{i_2} t^{a_3}_{i_3}$ ($i_1,i_2,i_3$ all different) in (\ref{ss20a}) yields
\begin{eqnarray}
\fl 0 =  \sum_{ k ( \neq i_4)} 2 \, w_{i_4 k} \, C^{a_1 a_2 a_3 a_4}_{i_1 i_2 i_3 k} +
\frac{1}{2} w_{i_4 i_1} \delta_{ a_1 a_4} C^{a_2 a_3}_{i_2 i_3} + \frac{1}{2} w_{i_4 i_2} \delta_{ a_2 a_4} C^{a_1 a_3}_{i_1 i_3} + \frac{1}{2} w_{i_4 i_3} \delta_{ a_3 a_4}
C^{a_1 a_2}_{i_1 i_2} \label{ss23} \\
+ w_{i_4 i_1} \left( \delta_{a_1 a_4} C^{b a_2 a_3 b}_{i_1 i_2 i_3 i_4} - C^{a_4 a_2 a_3 a_1}_{i_1 i_2 i_3 i_4} \right) +
w_{i_4 i_2} \left( \delta_{a_2 a_4} C^{a_1 b a_3 b}_{i_1 i_2 i_3 i_4} - C^{a_1 a_4 a_3 a_2}_{i_1 i_2 i_3 i_4} \right)\nonumber\\
+w_{i_4 i_3} \left( \delta_{a_3 a_4} C^{a_1 a_2 b b}_{i_1 i_2 i_3 i_4} - C^{a_1 a_2 a_4 a_3}_{i_1 i_2 i_3 i_4} \right),
\nonumber
\end{eqnarray}
where $C^{a_1 a_2}_{i_1 i_2} = \langle t^{a_1}_{i_1} t^{a_2}_{i_2} \rangle = \delta_{a_1 a_2} C^{(0)}_{i_1 i_2}=\delta_{a_1 a_2} C(|i_1-i_2|)/3$ is the two-point spin correlator found above.

The expression can be simplified noting that, by rotational invariance of the ground state, one has the decomposition
\begin{equation}
C^{a_1 a_2 a_3 a_4}_{i_1 i_2 i_3 i_4} =
\delta_{a_1 a_4} \delta_{a_2 a_3} C^{(1)}_{i_1 i_2 i_3 i_4} +
\delta_{a_2 a_4} \delta_{a_1 a_3} C^{(2)}_{i_1 i_2 i_3 i_4} +
\delta_{a_3 a_4} \delta_{a_1 a_2} C^{(3)}_{i_1 i_2 i_3 i_4},
 \label{ss22}
\end{equation}
where $C^{(a)}_{i_1 i_2 i_3 i_4} \; (a=1,2,3)$ are invariant tensors that can be obtained from $C^{a_1 a_2 a_3 a_4}_{i_1 i_2 i_3 i_4}$ as
\begin{eqnarray}
C^{(1)}_{i_1 i_2 i_3 i_4} =
\frac{1}{30} \left( 4 C^{a b b a}_{i_1 i_2 i_3 i_4} - C^{a b a b}_{i_1 i_2 i_3 i_4} - C^{a a b b }_{i_1 i_2 i_3 i_4} \right), \nonumber \\
C^{(2)}_{i_1 i_2 i_3 i_4} =
\frac{1}{30} \left( - C^{a b b a}_{i_1 i_2 i_3 i_4} + 4 C^{a b a b}_{i_1 i_2 i_3 i_4} - C^{a a b b }_{i_1 i_2 i_3 i_4} \right),  \label{ss22a} \\
C^{(3)}_{i_1 i_2 i_3 i_4} =
\frac{1}{30} \left( - C^{a b b a}_{i_1 i_2 i_3 i_4} - C^{a b a b}_{i_1 i_2 i_3 i_4} + 4 C^{a a b b }_{i_1 i_2 i_3 i_4} \right).
\nonumber
\end{eqnarray}
Permuting the spin matrices in (\ref{ss21}), one can easily relate these tensors,
\begin{eqnarray}
C_{i_1 i_2 i_3 i_4} & \equiv & C^{(1)}_{i_1 i_2 i_3 i_4} = C^{(2)}_{i_1 i_2 i_4 i_3} = C^{(3)}_{i_1 i_4 i_3 i_2}, \label{ss22b} \\
C_{i_1 i_2 i_3 i_4} &  = & C_{i_2 i_1 i_4 i_3} = C_{i_3 i_4 i_1 i_2} = C_{i_4 i_3 i_2 i_1}.\nonumber
\end{eqnarray}
Plugging (\ref{ss22}) into (\ref{ss23}), one finds an expression, where the terms proportional to $\delta_{a_1 a_4} \delta_{a_2 a_3}$, etc,
must cancel independently. This leads to
\begin{eqnarray}
\fl0=\sum_{ k ( \neq i_4)} 2 \, w_{i_4 k} \, C^{(1)}_{i_1 i_2 i_3 k}
+ 2 w_{i_4 i_1} C^{(1)}_{i_1 i_2 i_3 i_4}
+(w_{i_4 i_1} - w_{i_4 i_3} ) C^{(2)}_{i_1 i_2 i_3 i_4}\\
+ (w_{i_4 i_1} - w_{i_4 i_2} ) C^{(3)}_{i_1 i_2 i_3 i_4}
+ \frac{w_{i_4 i_1}}{2} C^{(0)}_{i_2 i_3},\nonumber\\
\fl0=\sum_{ k ( \neq i_4)} 2 \, w_{i_4 k} \, C^{(2)}_{i_1 i_2 i_3 k}
+ 2 w_{i_4 i_2} C^{(2)}_{i_1 i_2 i_3 i_4}
+(w_{i_4 i_2} - w_{i_4 i_1} ) C^{(3)}_{i_1 i_2 i_3 i_4}\\
+(w_{i_4 i_2} - w_{i_4 i_3} ) C^{(1)}_{i_1 i_2 i_3 i_4}
+ \frac{w_{i_4 i_2}}{2} C^{(0)}_{i_1 i_3},\nonumber\\
\fl0=\sum_{ k ( \neq i_4)} 2 \, w_{i_4 k} \, C^{(3)}_{i_1 i_2 i_3 k}
+ 2 w_{i_4 i_3} C^{(3)}_{i_1 i_2 i_3 i_4}  +
 (w_{i_4 i_3} - w_{i_4 i_2} ) C^{(1)}_{i_1 i_2 i_3 i_4}\\
+(w_{i_4 i_3} - w_{i_4 i_1} ) C^{(2)}_{i_1 i_2 i_3 i_4}
+\frac{w_{i_4 i_3}}{2} C^{(0)}_{i_1 i_2}.\nonumber
\end{eqnarray}
These equations can be solved numerically for the four-point correlators in terms of the two-point correlators which are known from the
solutions of (\ref{ss4}). We can also exploit the relations (\ref{ss22b}) to rewrite the above system in the compact form
\begin{eqnarray}\label{ss24}
\fl0=\sum_{ k ( \neq i_4)} 2 \, w_{i_4 k} \, C_{i_1 i_2 i_3 k}
+ 2 w_{i_4 i_1} C_{i_1 i_2 i_3 i_4}
+(w_{i_4 i_1} - w_{i_4 i_3} ) C_{i_1 i_2 i_4 i_3}\\
+ (w_{i_4 i_1} - w_{i_4 i_2} ) C_{i_1 i_4 i_3 i_2}
+ \frac{w_{i_4 i_1}}{2} C^{(0)}_{i_2 i_3}.\nonumber
\end{eqnarray}

\begin{figure}
\begin{indented}
\item[]\includegraphics[width=0.6\textwidth]{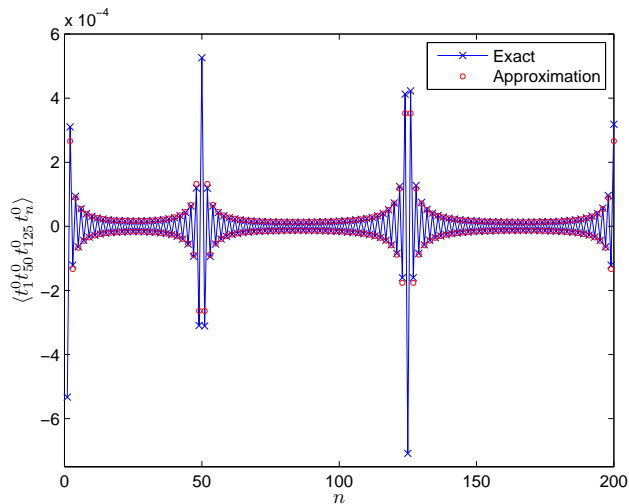}
\end{indented}
\caption{The four-point correlation function $\langle t_1^0t_{50}^0t_{125}^0t_n^0\rangle$ in the uniform model computed from the exact formula (\ref{ss24}) and the approximation (\ref{ss26}), which is valid when all the spins included in the four-point correlator are separated by many sites.}\label{fpcfig}
\end{figure}

For the uniform model, we can compare these formulas to the CFT prediction for the correlator of $M$ spins. We shall only consider the most relevant term in the expression (\ref{Si4}) for $t^0_n$, namely
\begin{eqnarray}\label{ss25}
t^0_n & \simeq &  (-1)^n \sqrt{ \frac{\pi}{2}} \cos \left( \frac{ \varphi(z_n, \bar{z}_n) }{\sqrt{2}} \right),
\end{eqnarray}
where $z_n = 2 \pi \rmi n \; (n=1, \dots, N)$. Employing the aforementioned map between the cylinder and the plane one can derive
\begin{eqnarray}\label{ss26}
\fl\langle t^0_{n_1} \dots t^0_{n_M} \rangle \simeq \left( \frac{\pi}{2} \right) ^{M/2} (-1)^{\sum_i n_i}
\left\langle \prod_{j= 1}^M \cos\left(\frac{ \varphi(z_{n_j}, \bar{z}_{n_j})  }{ \sqrt{2} }\right) \right\rangle_{\rm cyl} \\
= \left( \frac{ \pi}{ 16 N} \right)^{M/2} \, (-1)^{\sum_i n_i} \; \sum_{s_1, \dots, s_M = \pm 1} \delta_{ \sum_j s_j, 0}
\prod_{i < j}^M \left|  \sin\left[\frac{ \pi ( n_i - n_j)}{ N}\right] \right|^{ s_i s_j}. \nonumber
\end{eqnarray}
A comparison of this formula for $M=4$ and the results obtained from the numerical solution of (\ref{ss24}) in the uniform case is given in figure \ref{fpcfig}.

\section{Results for the $SU(2)$ WZW model at level $k=2$} \label{sec:keq2}

The $SU(2)_2$ WZW model has three primary fields $\phi_{j}$ associated to the spins $j=0$, $1/2$ and $1$. The nontrivial parts of the fusion rules are given by
\begin{equation}
\phi_{1/2} \otimes \phi_{1/2} = \phi_0 \oplus \phi_1, \qquad
\phi_1 \otimes \phi_1 = \phi_0, \qquad \phi_{1/2} \otimes \phi_1 = \phi_{1/2}.
\label{fu2}
\end{equation}
This theory is equivalent to the product of the critical Ising model and a free massless boson, so that their Virasoro central charges add up to $c = 1/2 + 1 = 3/2$. The current operators can be written in terms of the Majorana field $\chi(z)$ of the Ising model and vertex operators of the chiral boson as
\begin{equation}
J^0(z) = \rmi \, \partial_z \varphi(z), \qquad
J^\pm(z) = \sqrt{2} \; \chi(z) \, \rme^{ \pm \rmi \varphi(z)}.
\end{equation}
In the following, we investigate the model obtained from a product of $N$ spin 1 fields, and we note that the primary fields with spin $j=1$ and conformal weight $h_{1} = 1/2$ can be constructed as
\begin{equation}
\phi_{1,s_i}(z_i) =
\left\{ \begin{array}{cl}
 :\rme^{ \pm \rmi \varphi(z_i)}: & s_i=\pm1 \\
 \rho_{1,i}\chi(z_i) & s_i=0
 \end{array}
 \right.,
\qquad \rho_{1,i} \equiv
\left\{\begin{array}{cl}
 1 & i: {\rm even} \\
 -1 & i: {\rm odd}
 \end{array}
 \right..
\end{equation}
For brevity, we restrict the numerical analyses to the uniform case in this section. The Hamiltonian for the model obtained from a product of $N$ spin $1/2$ fields ($k=2$, $j=1/2$) is given in \ref{sec:k2spinhalf}.

\subsection{Wave function and Hamiltonian for $j=1$} \label{sec:Ham1}

The above representation of the primary fields enables us to compute the chiral correlator of $N$ spin 1 fields with spin projections $s_i = \pm 1, 0$ and parameters $z_i$ \cite{as10,cft-book}:
\begin{eqnarray}\label{psik2j1}
\fl\psi_{s_1, \dots, s_N}(z_1,\ldots,z_N) =
\rho_1\langle:\rme^{ \rmi s_1 \varphi(z_1)}:
\ldots:\rme^{ \rmi s_N \varphi(z_N)}:\rangle
\langle\chi(z_1)^{1-s_1^2}\ldots\chi(z_N)^{1-s_N^2}\rangle\nonumber\\
=\rho_1 \; \prod_{i < j}^N ( z_i - z_j)^{ s_i s_j} \; {\rm Pf}_0 \left( \frac{1}{z_i - z_j} \right),\\
\rho_1 \equiv (-1)^{ \sum_{i:{\rm odd}} (s_i -1)},\qquad
N:{\rm even}, \qquad \sum_{i=1}^N s_ i = 0.\nonumber
\end{eqnarray}
The Pfaffian ${\rm Pf}_0(\ldots)$ comes from the correlator of the $\chi(z)$ fields and is hence restricted to the positions where $s_i=0$. The sign factor $\rho_1$ ensures that $\psi$ is a singlet. The wave function vanishes if $N$ is odd because an odd number of $\phi_1$ fields fuse to a single $\phi_1$ field. We note that a different wave function containing some of the same elements was investigated in \cite{gt}, where a three-body Hamiltonian, whose ground state is close to the wave function, was found numerically.

To find the explicit expression for the Hamiltonian, we rewrite (\ref{cg16}) into
\begin{equation}
\left(K_{1,1}^{(i)}\right)_b^a=\frac{2}{3}\delta_{ab}
-\frac{5}{12}\rmi\vep_{abc}t_i^c-\frac{1}{12}(t_i^at_i^b+t_i^bt_i^a), \qquad
(t^a)_{bc}=-\rmi\vep_{abc},
\end{equation}
and consequently
\begin{equation}
C_{1,1}^{i,a}=\sum_{k(\neq i)}^N w_{ik}
\left[\frac{2}{3}t_k^a-\frac{5}{12}\rmi\vep_{abc}t_k^bt_i^c
-\frac{1}{12}(t_i^at_i^b+t_i^bt_i^a)t_k^b\right].
\end{equation}
When we plug this into (\ref{dec11}), assume $|z_i|=1$, $\forall i$, and exploit the relations
\begin{equation}
[t_i^a,t_i^b]=\rmi\vep_{abc}t_i^c, \qquad
t_i^at_i^a=2, \qquad
\vep_{abc}t_i^bt_i^c=\rmi t_i^a,
\end{equation}
we find
\begin{eqnarray}
\fl H_{1,1}^{(i)}=-\frac{4}{3}\sum_{j(\neq i)}^Nw_{ij}^2
-\frac{2}{3}\sum_{j\neq k(\neq i)}^Nw_{ij}w_{ik}t_j^at_k^a
-\frac{1}{3}\sum_{j(\neq i)}^Nw_{ij}^2t_i^at_j^a
+\frac{1}{6}\sum_{j(\neq i)}^Nw_{ij}^2(t_i^at_j^a)^2\\
+\frac{1}{6}\sum_{j\neq k(\neq i)}^Nw_{ij}w_{ik}t_i^at_j^at_i^bt_k^b.
\nonumber
\end{eqnarray}
The Hamiltonian for the spin 1 case is thus
\begin{eqnarray}\label{Hk2j1}
\fl H_{1,1}=-\frac{4}{3}\sum_{i\neq j}^Nw_{ij}^2
-\frac{1}{3}\sum_{i\neq j}^N\left(w_{ij}^2+2\sum_{k(\neq i,j)}^Nw_{ki}w_{kj}\right)t_i^at_j^a
+\frac{1}{6}\sum_{i\neq j}^Nw_{ij}^2(t_i^at_j^a)^2\\
+\frac{1}{6}\sum_{i\neq j\neq k}^Nw_{ij}w_{ik}t_i^at_j^at_i^bt_k^b.
\nonumber
\end{eqnarray}
The spectrum of (\ref{Hk2j1}) for the uniform case and $N=6$ is given in table \ref{SpectrumSpin1}. Since $H_{1,1}$ commutes with $T^a$, we again have that the $2S+1$ states within each spin multiplet are degenerate. Furthermore, we can also classify the states by the eigenvalue of the momentum operator $p_u=-\rmi \ln(\mathcal{T})$, where $\mathcal{T}$ is the translation operator defined in (\ref{trans}) with the permutation operator given in the spin 1 case by \cite{S41}
\begin{equation}
\mathcal{P}_{i,j}=(t^a_it^a_j)^2+t^a_it^a_j-1.
\end{equation}
We observe that all degeneracies between multiplets can be explained from the symmetry under translation and under a flip of the spin chain. This suggests that there is no Yangian symmetry in this model. We have tried to restore the Yangian symmetry by creating accidental degeneracies in the spectrum by giving the terms in the Hamiltonian (\ref{dec14}) different weights, but we did not find a solution.

\begin{table}
\lineup
\caption{\label{SpectrumSpin1}Spectrum of the spin 1 Hamiltonian (\ref{Hk2j1}) for $z_j=\exp(2\pi\rmi j/N)$ and $N=6$ (only the lowest 106 states are given). $E$ is the energy, $S$ is the spin, $p_u$ is the momentum, and $\#=2S+1$.}
\begin{indented}
\item[]\begin{tabular}{@{}llll}
\br
$E$ & $S$ & $\frac{3p_u}{\pi}$ & \# \\
\mr
   \00.0000&    0&    \ 0&  1\\
   \01.3322&    1&    \ 3&  3\\
   \03.8562&    2&    \ 0&  5\\
   \09.7852&    3&    \ 3&  7\\
   13.3457&    1&   -1&  3\\
          &    1&    \ 1&  3\\
   14.5984&    2&   -2&  5\\
          &    2&    \ 2&  5\\
\br
\end{tabular}
\hfill
\begin{tabular}{@{}llll}
\br
\multicolumn{4}{@{}l}{table continued\ldots} \\
\mr
   15.4088&    0&    \ 3&  1\\
   15.5224&    1&   -2&  3\\
          &    1&    \ 2&  3\\
   17.1898&    4&    \ 0&  9\\
   17.3510&    2&   -1&  5\\
          &    2&    \ 1&  5\\
   19.3269&    3&   -2&  7\\
          &    3&    \ 2&  7\\
\br
\end{tabular}
\hfill
\begin{tabular}{@{}llll}
\br
\multicolumn{4}{@{}l}{table continued\ldots} \\
\mr
   20.5618&    1&    \ 0&  3\\
   23.3377&    2&    \ 0&  5\\
   23.9121&    2&   -2&  5\\
          &    2&    \ 2&  5\\
   25.3391&    1&   -2&  3\\
          &    1&    \ 2&  3\\
   25.5448&    2&   -1&  5\\
          &    2&    \ 1&  5\\
\br
\end{tabular}
\end{indented}
\end{table}

\subsection{Comparison to the bilinear-biquadratic spin 1 chain} \label{sec:bilinear}

\begin{figure}
\begin{indented}
\item[]\includegraphics[width=0.6\textwidth]{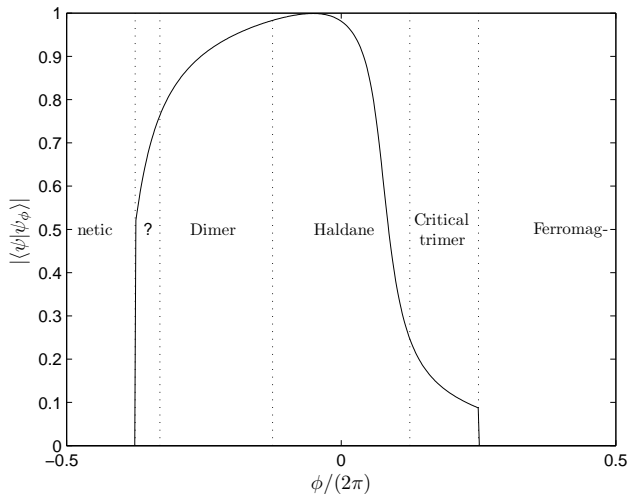}
\end{indented}
\caption{Overlap $|\langle\psi|\psi_\phi\rangle|$ between the ground state $\psi$ of (\ref{Hk2j1}) and the ground state $\psi_\phi$ of (\ref{Hblbq}) as a function of $\phi$ for $N=6$ sites (when the ground state of $H_\phi$ is degenerate, we choose the state in the ground state subspace, which maximizes $|\langle\psi|\psi_\phi\rangle|$). The dotted vertical lines indicate the transitions between different phases of the bilinear-biquadratic spin 1 chain in the thermodynamic limit.}\label{bilinbiquad}
\end{figure}

In \cite{CS10}, it was suggested that the dimer version of the generalized HS model for suitably chosen $\delta$ could be close to the spontaneously dimerized phase of the $J1$-$J2$ Heisenberg model because the Hamiltonians of both models involve terms of the form $t^a_it^a_{i+1}$ and $t^a_i t^a_{i+2}$ and because tuning $\delta$ can change the pattern of coupling strengths. It turned out that there indeed is a significant overlap between the wave functions of the two models, and one may ask whether the $k=2$ and $j=1$ model resembles another model. In the uniform case, a natural choice for comparison is the bilinear-biquadratic spin 1 Hamiltonian \cite{s87,sz93,owt}
\begin{equation}\label{Hblbq}
H_\phi=\sum_{i=1}^N\left[\cos(\phi)t_i^at_{i+1}^a
+\sin(\phi)\left(t_i^at_{i+1}^a\right)^2\right]
\end{equation}
with periodic boundary conditions $t^a_{N+1}=t^a_1$. We compare the state $\psi$ in (\ref{psik2j1}) to the ground state $\psi_\phi$ of (\ref{Hblbq}) for $N=6$ in figure \ref{bilinbiquad}. The phases of the bilinear-biquadratic spin 1 model for $N\rightarrow\infty$ are also indicated in the figure. In the Haldane phase, there is a unique, singlet ground state with an energy gap to the first exited state. For $N=6$, this is also what we have here, and hence one could expect that $\psi$ may be similar to $\psi_\phi$ in this parameter regime. In fact, the overlap between the two states is seen to be very close to unity in an interval within the Haldane phase. The maximum is $0.9990$ and occurs for $\phi = -0.3213$. A nonzero overlap is also seen in the dimer and trimer phases, where the condition $\sum_is_i=0$ is fulfilled. In the ferromagnetic phase, on the other hand, the overlap vanishes because the ground states of (\ref{Hblbq}) all have total spin $N$. The results presented below for the Renyi entropy and the two-point spin correlator suggest that the spin 1 model at level $k=2$ is gapless for $N\rightarrow\infty$, and hence we do not expect the two models to be close to each other in that limit.

\subsection{Entanglement entropy and two-point spin correlation function for $j=1$}\label{sec:entcor1}

It is interesting to ask, whether there is a simple connection between the properties of the conformal fields used to construct the wave function and the properties of the wave function. This turns out to be the case for the uniform HS model, where it is known that the low energy states are described by the $SU(2)_1$ WZW model in the large $N$ limit \cite{H88}. To investigate whether a similar result could be true for the spin 1 case, we compute below the Renyi entropy and the two-point correlator of the spin 1 chain numerically and compare the results to those expected for the $SU(2)_2$ WZW model.

\begin{figure}
\begin{indented}
\item[]\includegraphics[width=0.6\textwidth]{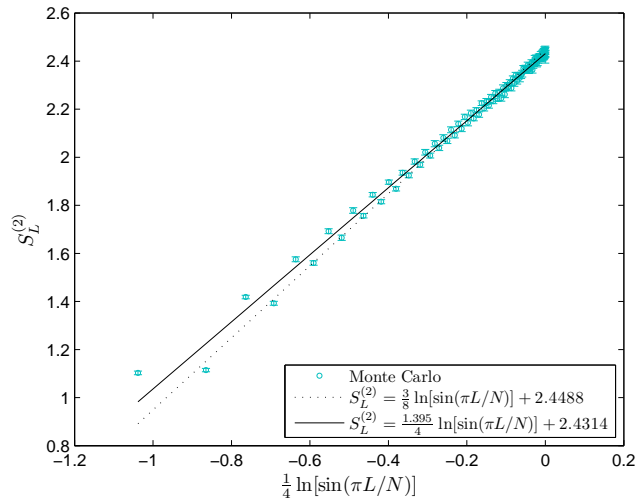}
\end{indented}
\caption{Renyi entropy of the state (\ref{psik2j1}) for $z_j=\exp(2\pi \rmi j/N)$ and $N=200$ obtained from Monte Carlo simulations of $\exp(-S_L^{(2)})$. The data shown is the average of 40 trajectories with different initial conditions, and the variation of the mean of these trajectories is used to estimate the error bars. The solid line is a fit of the form $S_L^{(2)}=c\ln(\sin(\pi L/N))/4+a$, where $a$ and $c$ are fitting parameters, and the dotted line is a fit of the same form but with $c$ fixed at $3/2$.}\label{S1entropy}
\end{figure}

We start with the Renyi entropy
\begin{equation}\label{SL}
S_L^{(2)}\equiv -\ln[\Tr(\rho_L^2)],
\end{equation}
where $\rho_L$ is the reduced density operator of the first $L$ spins obtained by tracing out the state of the last $N-L$ spins. Rewriting this equation, we can express the entropy in terms of an expectation value
\begin{eqnarray}\label{MCentropi}
\fl\rme^{-S_L^{(2)}}=
\sum_{s_1,\ldots,s_N}\sum_{s'_1,\ldots,s'_N}
|\psi_{s_1,\ldots,s_N}(z_1,\ldots,z_N)|^2|\psi_{s'_1,\ldots,s'_N}(z_1,\ldots,z_N)|^2\\
\times\frac{\psi^*_{s'_1,\ldots,s'_L,s_{L+1},\ldots,s_N}(z_1,\ldots,z_N)
\psi^*_{s_1,\ldots,s_L,s'_{L+1},\ldots,s'_N}(z_1,\ldots,z_N)}
{\psi^*_{s_1,\ldots,s_N}(z_1,\ldots,z_N)\psi^*_{s'_1,\ldots,s'_N}(z_1,\ldots,z_N)}\nonumber\\
\times\left[\sum_{s_1,\ldots,s_N}|\psi_{s_1,\ldots,s_N}(z_1,\ldots,z_N)|^2\right]^{-2},
\nonumber
\end{eqnarray}
which we evaluate numerically by use of the Metropolis Monte Carlo algorithm \cite{CS10}. The Renyi entropy of a conformal field theory with central charge $c$ is \cite{H94,L03,C04}
\begin{equation}\label{S2}
S_L^{(2)}=\frac{c}{4}\ln[\sin(\pi L/N)]+\textrm{constant},
\end{equation}
and $c=3/2$ for the $SU(2)_2$ WZW model (recall (\ref{su2})). In figure \ref{S1entropy}, we have plotted $S_L^{(2)}$ as a function of $\ln[\sin(\pi L/N)]/4$, and indeed the points fall almost on a straight line. Considering $c$ as a fitting parameter, we find that $c=1.395$ is the best choice to describe the data, but $c=3/2$ also gives a fit that approximately follow the data points.

\begin{figure}
\begin{indented}
\item[]\includegraphics[width=0.6\textwidth]{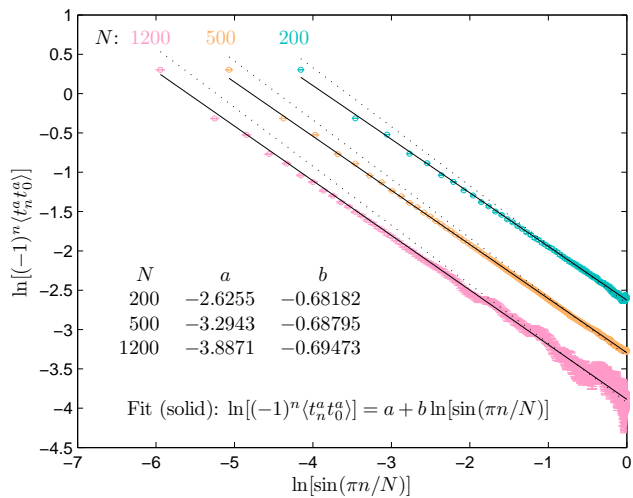}
\end{indented}
\caption{The functional dependence of the two-point spin correlation function on separation of the spins for $z_j=\exp(2\pi \rmi j/N)$ and $N=200$, $500$ and $1200$. The points are obtained from Monte Carlo simulations, the solid lines are fits of the form $\ln((-1)^n\langle t_n^at_0^a\rangle)=a+b\ln[\sin(\pi n/N)]$, where $a$ and $b$ are fitting parameters, and the dotted lines are fits of the same form but with $b$ fixed at $-3/4$. As in the previous figure, the error bars are estimated from the variation of the mean of trajectories with different initial conditions.}\label{figcor}
\end{figure}

Let us next consider the two-point spin correlation function. For the spin 1 model at level $k=2$, we do not obtain a closed set of equations for the two-point spin correlators from (\ref{ss20a}) and (\ref{ss20b}), so we use Monte Carlo simulations instead. Results for the uniform model and $N=200$, $500$ and $1200$ are shown in figure \ref{figcor}. Affleck found that the spin correlation function for an infinite chain of spin 1 fields in the WZW model at level $k=2$ is proportional to $(-1)^nn^{-3/4}$ for large $n$ \cite{a86}, possibly with a multiplicative logarithmic correction of the form $(A+B\ln(n))^{1/2}$, where $A$ and $B$ are constants \cite{AG89}. Referring back to (\ref{ss20}) and also to the scaling behaviour of the correlator in the dimer model discussed in section \ref{sec:dimer}, it is very natural to expect that the most significant term of the correlation function for the finite spin 1 chain is proportional to $(-1)^n[\sin(\pi n/N)]^b$ in the asymptotic region $n\gg1$, where the exponent $b$ may or may not be equal to $-3/4$. We therefore plot $\ln[(-1)^n\langle t_0^at_n^a\rangle]$ versus $\ln[\sin(\pi n/N)]$ in figure \ref{figcor}. This transformation indeed gives data lying almost on straight lines despite the fact that we are considering chains of finite length. Using a fit to extract the slope of the lines, we find that the exponent is around $-0.69$. A similar exponent was found for the state considered in \cite{ns04}. As in \cite{ns04}, we note that the deviation of $b$ from $-0.75$ may be due to the presence of a logarithmic correction since a logarithmic correction of the type given above is equivalent to a power law to first order in the exponent: $n^{\epsilon/2}\approx (1+\ln(n)\epsilon)^{1/2}$ for $\epsilon$ small. Altogether, it is hence possible that the properties of the wave function are those of the $SU(2)_2$ WZW model, but it may also be that there are some deviations.

\section{Conclusion}\label{sec:conclusion}

In conclusion, we have used a specific set of null vectors to derive quantum spin Hamiltonians, whose ground states are the chiral correlators of the $SU(2)_k$ WZW model, and we have shown that the same null vectors lead to a set of algebraic, linear equations relating the spin correlators within each model. At level $k=1$, these equations are sufficient to compute both two-point and higher-point spin correlators simply by solving a set of linear equations, which can even be done analytically in some cases. The models involve Hamiltonians with long-range interactions, and explicit analytical expressions for the corresponding ground state wave functions can be found in many cases. The examples presented in the paper include both new spin models and new results for already known models.

It is interesting to note that our construction in some sense resembles that of the AKLT \cite{AKLT,fnw} bond in matrix product states (MPS) and frustration free parent Hamiltonians. As in MPS, the components of the wave functions are constructed from a product of $N$ operators (see (\ref{dec1})), where the $j$th operator depends only on the state of the $j$th spin, but the operators are chiral conformal fields instead of finite dimensional matrices. Similarly, the Hamiltonian is a sum of $N$ Hamiltonians (see (\ref{dec14})) that annihilate the ground state individually, but the Hamiltonians involve long-range interactions rather than only local interactions. These differences, in fact, circumvent one of the limitations of MPS, namely how correlation functions may decay algebraically, and thus one can also use the states to describe critical behaviour.

The uniform Haldane-Shastry model is closely related to the Calogero-Sutherland (CS) model, whose Hamiltonian contains a kinetic term plus an inverse square coupling between particles moving on a circle \cite{GS} (see \cite{ThesisTalstra} for a review). In turn, particular eigenfunctions of the CS model are in direct correspondence with the correlation functions in boundary conformal field theory. In fact, the CS Hamiltonian can be obtained from the null vector decoupling equations at level 2 of boundary fields and a bulk primary field at the centre of the disk \cite{Cardy04,Cardy07}. These results suggest a possible relation of our work to those of \cite{Cardy04,Cardy07}, although the null vectors used there and those used in the present work are different.

The present paper raises the question, whether the properties of the wave functions are directly related to the properties of the fields used to construct the wave functions. The properties of the uniform model with $k=1$ and $j=1/2$ \cite{H88} and the numerical results presented above for the two-point spin correlator and the Renyi entropy for the uniform model with $k=2$ and $j=1$ suggest that the properties of the wave function may be similar to, but not necessarily identical to, the properties of the fields. We are currently investigating this question in more detail.

There are several ways in which the explicit examples given in the paper can be generalized. The construction in section \ref{sec:hamiltonians} already includes the possibility to consider models at higher levels, models in which the fields used to construct the wave functions do not all have the same spin index, and models in which the individual terms $H_{n_*, j}^{(i)}$ are given different weights in the final Hamiltonian. Furthermore, we have only considered one particular set of null vectors, but other Hamiltonians can be derived by starting from other null vectors. The present work shows that translationally invariant models of one-dimensional spin chains with long-range interactions can be achieved by choosing the $z_j$-coordinates to be uniformly distributed on the unit circle. More generally, the $z_j$-coordinates are numbers in the complex plane, and various interaction patterns can be achieved by adjusting these coordinates. This observation facilitates the construction of higher dimensional spin models. In particular, it is natural to describe a two-dimensional spin lattice in the above framework by interpreting, e.g., the real and imaginary parts of $z_j$ as the coordinates of the physical position of the $j$th spin in the lattice. This construction will be presented elsewhere \cite{ncs}. Finally, the idea of finding Hamiltonians from null vectors can also be applied to any Lie group and to WZW models based on supergroups. Altogether, the present work thus opens up possibilities to find a huge class of spin models. We conclude that null vectors are a powerful tool to gain insight into the behaviour of many-body systems, and we believe that more results will be obtained along these lines.

\vspace{5mm}

{\noindent\footnotesize\textit{Note added in proof.} A spin 1 model with the same type of interactions as the model discussed in section 4 has
been derived for the uniform case in \cite{Greiter}, and an application of conformal field theory to propose wave functions for the fractional quantum Hall effect is considered in \cite{mr91}.}

\ack
The authors acknowledge discussions with D. Haldane, D. Doyon, G. Mussardo, D. Bernard, M. B. Hastings, V. Gurarie, C. Nayak and H.-H. Tu. This work has been supported by The Carlsberg Foundation, the EU project QUEVADIS, and the grants FIS2009-11654 and QUITEMAD.

\appendix

\section{Alternative derivation of the generalized HS model}
\label{sec:alternative}

In \cite{CS10}, it was found by use of the Knizhnik-Zamolodchikov equation that the state (\ref{psi2}) is an eigenstate of the generalized HS Hamiltonian (\ref{hs3}). We provide the details of this derivation in the following. We start from the KZ equation
\begin{equation}
\frac{3}{2} \frac{\partial}{ \partial z_i} \; \psi =
\sum_{j ( \neq i)} \frac{ t^a_i t^a_j}{ z_i - z_j}  \; \psi,
\qquad i=1, \dots, N,
\end{equation}
for the chiral correlator (\ref{psi1}) with components (\ref{psi2}). Using the conformal transformation $z=\rme^w$, one can map the complex plane into the cylinder, where the chiral correlator becomes
\begin{equation}
\psi_{\rm cyl}(w_1, \dots, w_N) = \left( \frac{ \rmd w_1}{\rmd z_1} \right)^{-h} \dots \left( \frac{ \rmd w_N}{\rmd z_N } \right)^{-h}
\; \psi_{\rm plane}(z_1, \dots, z_N)
\end{equation}
and $h = 1/4$ is the conformal weight of the spin 1/2 primary field.
Expressing $\psi_{\rm cyl}$ in terms of the $z_i$ variables and using (\ref{psi2}) for $\psi_{\rm plane}(z_1, \dots, z_N)\equiv\psi(z_1, \dots, z_N)$, one has
\begin{equation}\label{psicyl}
\fl\psi_{\rm cyl}(z_1, \dots, z_N) = \prod_i z_i^{1/4} \; \psi_{\rm plane}(z_1, \dots, z_N)
= \rho_{\frac{1}{2}} \; \prod_{ i < j} (z_i - z_j)^{s_i s_j/2} \prod_i z_i^{1/4}.
\end{equation}
The KZ equation for this wave function is
\begin{equation}\label{KZ}
3 z_i \frac{\partial}{ \partial z_i} \; \psi_{\rm cyl} =
\sum_{j ( \neq i)} w_{ij} \; t^a_it^a_j \; \psi_{\rm cyl},
\qquad i=1, \dots, N,
\end{equation}
where $w_{ij}=(z_i+z_j)/(z_i-z_j)$ as usually.

Using the explicit expression for the wave function in (\ref{psicyl}) and the relations
\begin{equation}
\sum_{ j ( \neq i)} s_i s_j = - s_i^2 = -1, \qquad \sum_j s_j = 0,
\end{equation}
one finds
\begin{equation}
\fl z_i \frac{\partial}{ \partial z_i} \; \psi_{\rm cyl} = \left( \frac{1}{2}
\sum_{j ( \neq i)} \frac{ z_i}{z_i - z_j} s_i s_j + \frac{1}{4} \right) \psi_{\rm cyl}
=\frac{1}{2}\sum_{j ( \neq i)} \left( \frac{ z_i}{z_i - z_j} - \frac{1}{2} \frac{ z_i- z_j}{z_i - z_j} \right) s_i s_j \psi_{\rm cyl}.
\end{equation}
Simplifying the above expression one gets an abelian version of the KZ equation
\begin{equation}\label{aKZ}
z_i \frac{\partial}{ \partial z_i} \; \psi_{\rm cyl} = \frac{1}{4}
 \sum_{j ( \neq i)}w_{i j} \; s_i s_j \; \psi_{\rm cyl}.
\end{equation}
Taking another derivative, one finds
\begin{equation}
\left( z_i \frac{\partial}{ \partial z_i} \right)^2 \psi_{\rm cyl} = - \frac{1}{2}
 \sum_{j ( \neq i)} \frac{ z_i z_j}{ z_{ij}^2} \; s_i s_j \; \psi_{\rm cyl} +
  \frac{1}{16}
 \sum_{j ( \neq i)}  \sum_{k ( \neq i)}
 w_{i j} w_{ik}  \; s_j s_k \; \psi_{\rm cyl},
\end{equation}
where we have used $s_i^2 = 1$ and
\begin{equation}
z_i \frac{\partial}{ \partial z_i} w_{ij} = - \frac{ 2 z_i z_j}{z_{ij}^2}.
\end{equation}
Summing over $i$ in the last expression, one gets
\begin{eqnarray}
\Delta \psi_{\rm cyl} & \equiv \sum_i \left( z_i \frac{\partial}{ \partial z_i} \right)^2 \psi_{\rm cyl}\\
& = - \frac{1}{2}\sum_{i \neq j} \frac{ z_i z_j}{ z_{ij}^2} \; s_i s_j \; \psi_{\rm cyl} +\frac{1}{16} \sum_i \sum_{j ( \neq i)}  \sum_{k ( \neq i)}
 w_{i j} w_{ik}  \; s_j s_k \; \psi_{\rm cyl}.\nonumber
\end{eqnarray}
The triple sum can be written as
\begin{equation}
\sum_i\sum_{j ( \neq i)}  \sum_{k ( \neq i)} = \sum_{i \neq j} \delta_{j k} +
\sum_{j \neq k}  \sum_{i ( \neq j, k)}
\end{equation}
so that
\begin{equation}
\fl\Delta \psi_{\rm cyl} = - \frac{1}{2}
\sum_{i \neq j} \frac{ z_i z_j}{ z_{ij}^2} \; s_i s_j \; \psi_{\rm cyl} +
\frac{1}{16}\sum_{i \neq j} w_{ij}^2  \; \psi_{\rm cyl} +
\frac{1}{16}\sum_{j \neq k}  \sum_{ i ( \neq j, k )}
w_{i j} w_{ik}  \; s_j s_k \; \psi_{\rm cyl}.
\end{equation}
Exchanging $ i \leftrightarrow k$, this becomes
\begin{equation}\label{dpsi}
\fl\Delta \psi_{\rm cyl} = - \frac{1}{2}
\sum_{i \neq j} \frac{ z_i z_j}{ z_{ij}^2} \; s_i s_j \; \psi_{\rm cyl} +
\frac{1}{16}\sum_{i \neq j} w_{ij}^2  \; \psi_{\rm cyl} +
\frac{1}{16}\sum_{i \neq j} s_i s_j  \sum_{ k ( \neq i,j )}
w_{k i } w_{k j} \; \psi_{\rm cyl}.
\end{equation}
Using (\ref{wid1}), (\ref{wid2}) and $\sum_{ i \neq j} s_i s_j = - N$ in (\ref{dpsi}), we find
\begin{eqnarray}
\fl\frac{\Delta \psi_{\rm cyl}}{\psi_{\rm cyl}} = - \frac{1}{2}
 \sum_{i \neq j} \frac{ z_i z_j}{ z_{ij}^2} \; s_i s_j +
  \frac{1}{16}
 \sum_{i \neq j} w_{ij}^2  +
  \frac{1}{16}
 \sum_{i \neq j} s_i s_j
[ N-2 + 2 w_{ij}^2 + w_{ij} (c_i - c_j) ] \\
= - \frac{1}{2}
 \sum_{i \neq j} \left( \frac{ z_i z_j}{ z_{ij}^2} - \frac{1}{4} w_{ij}^2 \right) \; s_i s_j  +
  \frac{1}{16}  \sum_{i \neq j} w_{ij}^2  +  \frac{N-2}{16}
 \sum_{i \neq j} s_i s_j\nonumber\\
 +\frac{1}{16}
 \sum_{i \neq j} s_i s_j \; w_{ij} (c_i - c_j)
=  \frac{1}{8}
 \sum_{i \neq j} s_i s_j +  \frac{N-2}{16}
 \sum_{i \neq j} s_i s_j + \frac{1}{16}  \sum_{i \neq j} w_{ij}^2 \nonumber\\
 +\frac{1}{16}
 \sum_{i \neq j} s_i s_j  w_{ij} (c_i - c_j)
= - \frac{N^2}{16}
 + \frac{1}{16}  \sum_{i \neq j} w_{ij}^2 +
  \frac{1}{16}
 \sum_{i \neq j} s_i s_j  w_{ij} (c_i - c_j).  \nonumber
\end{eqnarray}
From (\ref{aKZ}), we have
\begin{equation}
\frac{1}{2} \sum_i c_i z_i \frac{\partial}{ \partial z_i} \; \psi_{\rm cyl} =  \frac{1}{16}
 \sum_{i \neq j} s_i s_j  w_{ij} (c_i - c_j) \psi_{\rm cyl},
\end{equation}
so finally
\begin{equation}\label{LHS1}
 \left( \Delta -
\frac{1}{2} \sum_i c_i z_i \frac{\partial}{ \partial z_i} \right) \; \psi_{\rm cyl} =
\left( - \frac{N^2}{16}
 + \frac{1}{16}  \sum_{i \neq j} w_{ij}^2 \right) \psi_{\rm cyl}.
\end{equation}

The operator on the left hand side of (\ref{LHS1}) can also be computed from the KZ equation (\ref{KZ}). Taking the derivative of (\ref{KZ}) and following the same steps as before, one finds
\begin{equation}
\fl 9 \; \Delta \psi_{\rm cyl} = - 6 \sum_{ i \neq j}
\frac{z_i z_j}{z_{ij}^2} t_i^at_j^a \psi_{\rm cyl}+
\sum_{i \neq j} w_{ij}^2 (t_i^at_j^a)^2 \psi_{\rm cyl} +
\sum_{i \neq j} \sum_{k (\neq i, j)} w_{ki} w_{kj} t_k^at_i^at_k^bt_j^b  \psi_{\rm cyl}.
\end{equation}
The following identities are needed
\begin{eqnarray}
( t_i^at_j^a )^2 = \frac{3}{16} - \frac{1}{2} t_i^at_j^a , \qquad i \neq j,\\
t_k^at_i^at_k^bt_j^b =
 \frac{1}{4} t_i^at_j^a + \frac{\rmi}{2} \; \vep_{abc}t_k^at_i^bt_j^c, \qquad i \neq j \neq k,\\
 \sum_{i \neq j} t_i^at_j^a = T^aT^a - \frac{3}{4} N,
\end{eqnarray}
which finally lead to
\begin{equation*}
\fl\Delta \psi_{\rm cyl} = \left[ - \frac{2}{3} \sum_{ i \neq j}
\frac{z_i z_j}{z_{ij}^2} t_i^at_j^a + \frac{1}{48}
\sum_{i \neq j} w_{ij}^2  - \frac{N (N-2)}{48}  +
 \frac{1}{36} \sum_{i \neq j}
w_{ij} ( c_i - c_j)t_i^at_j^a\right]\psi_{\rm cyl},
\end{equation*}
where we have used $T^aT^a\psi_{\rm cyl}=0$ because $\psi_{\rm cyl}$ is a singlet. It also follows from (\ref{KZ}) that
\begin{equation}
\frac{1}{2} \sum_i c_i z_i \frac{\partial}{ \partial z_i} \psi_{\rm cyl}
= \frac{1}{12} \sum_{i \neq j} w_{ij} ( c_i - c_j)
t_i^at_j^a\psi_{\rm cyl},
\end{equation}
and hence
\begin{eqnarray}\label{LHS2}
\fl\left( \Delta - \frac{1}{2} \sum_i c_i z_i \frac{\partial}{ \partial z_i} \right)
\psi_{\rm cyl} = \Bigg\{ - \frac{2}{3} \sum_{ i \neq j}
\left[ \frac{z_i z_j}{z_{ij}^2} + \frac{1}{12} w_{ij} ( c_i - c_j) \right]
t_i^at_j^a + \frac{1}{48}\sum_{i \neq j} w_{ij}^2\\
- \frac{N (N-2)}{48}  \Bigg\}  \psi_{\rm cyl}.\nonumber
\end{eqnarray}
Equating (\ref{LHS1}) and (\ref{LHS2}), we arrive at
\begin{equation}\label{HScs}
\fl- \sum_{ i \neq j}
\left[ \frac{z_i z_j}{z_{ij}^2} + \frac{1}{12} w_{ij} ( c_i - c_j) \right]
t_i^at_j^a \; \psi_{\rm cyl} =
\left[ \frac{1}{16}
\sum_{i \neq j} w_{ij}^2  - \frac{N (N+1)}{16}  \right]  \psi_{\rm cyl}.
\end{equation}
Since $\psi_{\rm cyl}$ only differs from $\psi$ by an $s_i$-independent factor, we can replace $\psi_{\rm cyl}$ by $\psi$ in (\ref{HScs}), and it follows that $\psi$ is an eigenstate of (\ref{hs3}) as claimed. In \cite{CS10}, it was found numerically for small systems that $\psi$ is, in fact, the ground state of (\ref{hs3}). With the results of the present paper, however, it is clear from the construction that $\psi$ is the ground state. We note also that the above derivation is not immediately generalizable to $k>1$.

\subsection{The Inozemtsev invariant}

Let us finally use the above approach to derive the generalization (\ref{H3}) to the nonuniform case of the Inozemtsev invariant. We start from the second derivative
\begin{equation}
 \left( z_i \frac{\partial}{ \partial z_i} \right)^2 \psi_{\rm cyl} = - \frac{1}{2}
 \sum_{j ( \neq i)} \frac{ z_i z_j}{ z_{ij}^2} \; s_i s_j \; \psi_{\rm cyl} +
  \frac{1}{16}
 \sum_{j ( \neq i)}  \sum_{k ( \neq i)}
 w_{i j} w_{ik}  \; s_j s_k \; \psi_{\rm cyl}
\end{equation}
and take another derivative
\begin{eqnarray}
\fl\frac{\Delta_3 \psi_{\rm cyl}}{ \psi_{\rm cyl}} = \frac{1}{\psi_{\rm cyl}} \sum_i \left( z_i \frac{\partial}{ \partial z_i} \right)^3 \psi_{\rm cyl}  = - \frac{1}{2}\sum_{i \neq j} z_i \frac{ \partial }{ \partial z_i } \left( \frac{ z_i z_j}{ z_{ij}^2} \right)  \; s_i s_j\\
- \frac{1}{8}\sum_{i \neq j}  \sum_{ k ( \neq i)} \frac{ z_i z_j}{ z_{ij}^2}\; w_{ik}  s_j s_k
+ \frac{1}{16} \sum_i  \sum_{j ( \neq i)}  \sum_{k ( \neq i)}
z_i \frac{ \partial }{ \partial z_i } \left( w_{i j} w_{ik} \right) s_j s_k \nonumber\\
+\frac{1}{64} \sum_i  \sum_{j ( \neq i)}  \sum_{k ( \neq i)}  \sum_{l ( \neq i)}w_{i j} w_{ik}  w_{il}  \; s_i s_j s_k s_l.
\nonumber
\end{eqnarray}
The terms involving 4 different spin variables vanish by virtue of the equation
\begin{equation}
 w_{i j} w_{ik}  w_{il} + (\textrm{permutations of } i,j,k,l) = 0, \qquad i\neq j \neq k \neq l.
\end{equation}
Then, using the following identities
\begin{eqnarray}
z_i \frac{\partial}{ \partial z_i} w_{ij} = - \frac{ 2 z_i z_j}{z_{ij}^2}, \qquad
z_i \frac{\partial}{ \partial z_i}\frac{ z_i z_j}{z_{ij}^2} = \frac{ 1}{4}
( w_{ij} - w^3_{ij}), \\
z_k \frac{\partial}{ \partial z_k} ( w_{ki } w_{kj} ) = \frac{1}{2} ( w_{ki} - w_{ki} w^3_{kj} + w_{kj} - w_{kj} w^2_{ki}),
\end{eqnarray}
we find after several manipulations
\begin{equation}
\Delta_3 \psi_{\rm cyl} = - \frac{3}{64} \sum_{i \neq j} c_{i,2} \; w_{ij} s_i s_j \psi_{\rm cyl}, \qquad c_{i,2} \equiv \sum_{k (\neq i)} w_{ki}^2.
\end{equation}
Hence,
\begin{equation}
\left( \Delta_3 + \frac{3}{16} \sum_i c_{i,2} \; z_i \frac{ \partial }{ \partial z_i} \right) \psi_{\rm cyl} =0.
\end{equation}
The left hand side can be computed using the KZ equation. After a long computation one gets $H_3\psi_{\rm cyl}=0$ and hence $H_3\psi=0$, where $H_3$ is given by (\ref{H3}).

\section{Hamiltonian for $k=2$ and $j=1/2$}\label{sec:k2spinhalf}

The Hamiltonian at level $k=2$, which has the chiral correlator $\langle\phi_{1/2, s_1} (z_1)\ldots\phi_{1/2, s_N} (z_N)\rangle$ of $N$ spin $1/2$ fields as its ground state, follows from (\ref{cg15}), (\ref{dec10}), (\ref{dec11}) and (\ref{dec14}). Assuming $|z_i|=1$ $\forall i$, we find
\begin{eqnarray}
\fl H_{2,1/2}^{(i)}=\frac{3}{8}\sum_{i_1\neq i_2(\neq i)}w_{ii_1}^2w_{ii_2}^2
+\sum_{i_1\neq i_2\neq i_3(\neq i)}
w_{ii_1}w_{ii_2}w_{ii_3}^2t_{i_1}^at_{i_2}^a
+\sum_{i_1\neq i_2(\neq i)}
w_{ii_1}^2w_{ii_2}^2t_i^at_{i_1}^a\\
+\frac{1}{2}\sum_{i_1\neq i_2(\neq i)}
w_{ii_1}^2w_{ii_2}^2t_{i_1}^at_{i_2}^a
+\frac{2}{5}\sum_{i_1\neq i_2\neq i_3\neq i_4(\neq i)}
w_{ii_1}w_{ii_2}w_{ii_3}w_{ii_4}t_{i_1}^at_{i_2}^at_{i_3}^bt_{i_4}^b\nonumber\\
+\frac{2}{5}\sum_{i_1\neq i_2\neq i_3(\neq i)}
w_{ii_1}w_{ii_2}w_{ii_3}(4w_{ii_1}-w_{ii_2}-w_{ii_3})
t_i^at_{i_1}^at_{i_2}^bt_{i_3}^b\nonumber
\end{eqnarray}
and
\begin{eqnarray}\label{H212}
\fl H_{2,1/2}=\frac{3}{8}\sum_{i\neq i_1\neq i_2}
w_{ii_1}^2w_{ii_2}^2
+\sum_{i_1\neq i_2\neq i_3}
\Bigg(\sum_{i(\neq i_1,i_2,i_3)}
w_{ii_1}w_{ii_2}w_{ii_3}^2+w_{i_1i_2}^2w_{i_1i_3}^2\\
+\frac{1}{2}w_{i_1i_3}^2w_{i_2i_3}^2\Bigg)t_{i_1}^at_{i_2}^a
+\frac{2}{5}\sum_{i_1\neq i_2\neq i_3\neq i_4} \Bigg[\sum_{i(\neq i_1,i_2,i_3,i_4)}
w_{ii_1}w_{ii_2}w_{ii_3}w_{ii_4}\nonumber\\
+w_{i_1i_2}w_{i_1i_3}w_{i_1i_4}(4w_{i_1i_2}-w_{i_1i_3}-w_{i_1i_4})\Bigg]
t_{i_1}^at_{i_2}^at_{i_3}^bt_{i_4}^b.\nonumber
\end{eqnarray}
The spectrum of (\ref{H212}) for the uniform case and $N=6$ is given in table \ref{spectraonehalf}. Since two $\phi_{1/2}$ fields fuse to a $\phi_0$ and a $\phi_1$ field, and since the fields should altogether fuse to $\phi_0$ to give a nonzero correlator, the ground state is $2^{N/2-1}$ times degenerate. Apart from this degeneracy, all degeneracies between spin multiplets can be explained from the invariance of the chain under translation by one site and under a flip of the chain. This suggests that there is no Yangian symmetry at level $k=2$. Another difference compared to the uniform HS model is that not all energies are integers.

\begin{table}
\lineup
\caption{\label{spectraonehalf}Spectrum of the spin 1/2 Hamiltonian (\ref{H212}) for $z_j=\exp(2\pi\rmi j/N)$ and $N=6$ sites. $E$ is the energy, $S$ is the spin, $p_u$ is the momentum, and $\#=2S+1$.}
\begin{indented}
\item[]\begin{tabular}{@{}llll}
\br
$E$ & $S$ & $\frac{3p_u}{\pi}$ & \# \\
\mr
    \0\00.0000&    0&   -1&  1\\
    &    0&  \ 0&  1\\
    &    0&  \ 1&  1\\
    &    0&  \ 3&  1\\
    \0\00.2715&    1&  \ 0&  3\\
    \0\02.8686&    1&   -2&  3\\
    &    1&  \ 2&  3\\
\br
\end{tabular}
\hfill
\begin{tabular}{@{}llll}
\br
\multicolumn{4}{@{}l}{table continued\ldots} \\
\mr
   \020.0000&    1&   -1&  3\\
   &    1&  \ 1&  3\\
   \034.4000&    2&   -2&  5\\
   &    2&  \ 2&  5\\
   \047.2000&    2&  \ 3&  5\\
   \052.0000&    1&  \ 3&  3\\
   \058.9285&    1&  \ 0&  3\\
\br
\end{tabular}
\hfill
\begin{tabular}{@{}llll}
\br
\multicolumn{4}{@{}l}{table continued\ldots} \\
\mr
  107.2000&    3&  \ 0&  7\\
  111.2000&    2&   -1&  5\\
  &    2&  \ 1&  5\\
  117.1314&    1&   -2&  3\\
  &    1&  \ 2&  3\\
  160.0000&    0&  \ 3&  1\\
  \\
\br
\end{tabular}
\end{indented}
\end{table}


\section*{References}

\begin{thebibliography}{99}

\bibitem{Bethe} Bethe H, {\it Zur Theorie der Metalle, I. Eigenwerte und Eigenfunktionen der linearen Atomkette}, 1931 {\it Zeitschrift f\"ur Physik} A {\bf 71} 205

\bibitem{AKLT} Affleck I, Kennedy T, Lieb E H and Tasaki H, {\it Rigorous results on valence-bond ground states in antiferromagnets}, 1987 \PRL {\bf 59} 799

\bibitem{H88} Haldane F D M, {\it Exact Jastrow-Gutzwiller resonating-valence-bond ground state of the spin-$\frac{1}{2}$ antiferromagnetic Heisenberg chain with $1/r^2$ exchange}, 1988 \PRL {\bf 60} 635

\bibitem{S88} Shastry B S, {\it Exact solution of an S=1/2 Heisenberg antiferromagnetic chain with long-ranged interactions}, 1988 \PRL {\bf 60} 639

\bibitem{as10} Ardonne E and Sierra G, {\it Chiral correlators of the Ising conformal field theory}, 2010 {\it J. Phys. A: Math. Theor.} {\bf43} 505402

\bibitem{CS10} Cirac J I and Sierra G, {\it Infinite matrix product states, conformal field theory, and the Haldane-Shastry model}, 2010 \PR B {\bf 81} 104431 ({\it Preprint} arXiv:0911.3029)

\bibitem{KZ84} Knizhnik V G and Zamolodchikov A B, {\it Current algebra and Wess-Zumino model in two dimensions}, 1984 \NP B {\bf247} 83

\bibitem{GW} Gepner D and Witten E, {\it String theory on group manifolds}, 1986 \NP B {\bf278} 493

\bibitem{cft-book} Di Francesco P, Mathieu P and S\'en\'echal D, 1997 {\it Conformal Field Theory} (New York: Springer-Verlag)

\bibitem{ThesisTalstra} Talstra J C, {\it Integrability and applications of the exactly-solvable Haldane-Shastry one-dimensional quantum spin chain}, 1995 {\it PhD thesis} Princeton University {\it Preprint} arXiv:cond-mat/9509178

\bibitem{H91} Haldane F D M, {\it ``Spinon gas'' description of the S=1/2 Heisenberg chain with inverse-square exchange: Exact spectrum and thermodynamics}, 1991 \PRL {\bf 66} 1529

\bibitem{yang} Haldane F D M, Ha Z N C, Talstra J C, Bernard D and Pasquier V, {\it Yangian symmetry of integrable quantum chains with long-range interactions and a new description of states in conformal field theory}, 1992 \PRL {\bf 69} 2021

\bibitem{D85} Drinfeld V G, {\it Hopf algebras and the quantum Yang-Baxter equation} 1985 {\it Sov. Math. Dokl.} {\bf 32} 254

\bibitem{I90} Inozemtsev V I, {\it On the connection between the one-dimensional $S=1/2$ Heisenberg chain and Haldane-Shastry model} 1990 {\it J. Stat. Phys.} {\bf 59} 1143

\bibitem{RM04} Refael G and Moore J E, {\it Entanglement entropy of random quantum critical points in one dimension}, 2004 \PRL {\bf93} 260602

\bibitem{RM09} Refael G and Moore J E, {\it Criticality and entanglement in random quantum systems}, 2009 {\it J. Phys. A: Math. Theor.} {\bf42} 504010

\bibitem{GV87} Gebhard F and Vollhardt D, {\it Correlation functions for Hubbard-type models: The exact results for the Gutzwiller wave function in one dimension}, 1987 \PRL {\bf 59} 1472

\bibitem{AH87} Affleck I and Haldane F D M, {\it Critical theory of quantum spin chains}, 1987 \PR B {\bf 36} 5291

\bibitem{A88} Affeck I, Les Houches Lecture Notes, in: Fields, Strings, and Critical Phenomena, ed. Brezin E and Zinn-Justin J (North-Holland, Amsterdam, 1988)

\bibitem{AG89} Affleck I, Gepner D, Schulz H J and Ziman T, {\it Critical behaviour of spin-s Heisenberg antiferromagnetic chains: analytic and numerical results}, 1989 \JPA {\bf 22} 511

\bibitem{E96} Eggert S, {\it Numerical evidence for multiplicative logarithmic corrections from marginal operators}, 1996 \PR B {\bf 54} R9612

\bibitem{dpc05} Deng X-L, Porras D and Cirac J I, {\it Effective spin quantum phases in systems of trapped ions}, 2005 \PR A {\bf 72} 063407

\bibitem{scw05} Schuch N, Cirac J I and Wolf M M, {\it Quantum States on Harmonic Lattices}, 2006 {\it Commun. Math. Phys.} {\bf267} 65 ({\it Preprint} arXiv:quant-ph/0509166)

\bibitem{gt} Greiter M and Thomale R, {\it Non-Abelian Statistics in a Quantum Antiferromagnet}, 2009 \PRL {\bf102} 207203

\bibitem{S41} Schr\"odinger E, {\it Exchange and spin}, 1941 {\it Proc. R. Ir. Acad. Sect.} A {\bf47} 39

\bibitem{s87} S\'olyom J, {\it Competing bilinear and biquadratic exchange couplings in spin-1 Heisenberg chains}, 1987 \PR B {\bf 36} 8642

\bibitem{sz93} Schadschneider A and Zittartz J, {\it Variational study of isotropic spin-1 chains using matrix-product states}, 1995 {\it Annalen der Physik} {\bf507} 157

\bibitem{owt} Orus R, Wei T-C and Tu H-H, {\it Phase diagram of the SO(n) bilinear-biquadratic chain from many-body entanglement}, 2011 \PR B {\bf 84} 064409

\bibitem{H94} Holzhey C, Larsen F and Wilczek F, {\it Geometric and renormalized entropy in conformal field theory}, 1994 \NP B {\bf 424} 443

\bibitem{L03} Vidal G, Latorre J I, Rico E and Kitaev A, {\it Entanglement in quantum critical phenomena}, 2003 \PRL {\bf 90} 227902

\bibitem{C04} Calabrese P and Cardy J, {\it Entanglement entropy and quantum field theory}, 2004 {\it J. Stat. Mech.} P06002

\bibitem{a86} Affleck I, {\it Exact critical exponents for quantum spin chains, non-linear $\sigma$-models at $\theta=\pi$ and the quantum hall effect}, 1986 \NP B {\bf 265} 409

\bibitem{ns04} Narayan O and Shastry B S, {\it Spin-s wave functions with algebraic order}, 2004 \PR B {\bf70} 184440

\bibitem{GS} Sutherland B, 2004 {\it Beautiful Models: 70 Years of Exactly Solved Quantum Many-Body Problems} (Singapore: World Scientific)

\bibitem{Cardy04} Cardy J, {\it Calogero-Sutherland model and bulk-boundary correlations in conformal field theory}, 2004 \PL B {\bf582} 121 ({\it Preprint} arXiv:hep-th/0310291)

\bibitem{Cardy07} Doyon B and Cardy J, {\it Calogero-Sutherland eigenfunctions with mixed boundary conditions and conformal field theory correlators}, 2007 {\it J. Phys. A: Math. Theor.} {\bf40} 2509 ({\it Preprint} arXiv:hep-th/0611054)

\bibitem{fnw} Fannes M, Nachtergaele B and Werner R F, {\it Finitely Correlated States on Quantum Spin Chains}, 1992 {\it Commun. Math. Phys.} {\bf 144} 443

\bibitem{ncs} Nielsen A E B, Cirac J I and Sierra G, in preparation.

\bibitem{Greiter} Greiter M, 2011 {\it Mapping of Parent Hamiltonians, From Abelian and non-Abelian Quantum Hall States to Exact Models of Critical Spin Chains} (\textit{Springer Tracts in Modern Physics} vol 244) (Berlin: Springer)

\bibitem{mr91} Moore G and Read N, {\it Nonabelions in the fractional quantum Hall effect}, 1991 \NP B {\bf 360} 362

\end{thebibliography}
\end{document}